\documentclass[useAMS,referee,usenatbib]{arxiv_biom}

\usepackage{graphicx}
\usepackage{bbm}
\usepackage{xcolor}
\usepackage{url}
\usepackage{appendix}
\usepackage{algorithm}
\usepackage{algorithmic}
\usepackage{amsmath,amssymb}
\usepackage{subfigure}
\usepackage{booktabs}
\usepackage{multirow}
\allowdisplaybreaks
\def\bSig\mathbf{\Sigma}

\title[Estimating causal excursion effects in mobile health with zero-inflated count outcomes]{Incorporating nonparametric methods for estimating causal excursion effects in mobile health with zero-inflated count outcomes}

 \author{
 	Xueqing Liu$^{1}$, Tianchen Qian$^{2}$, Lauren Bell$^{3,4}$,  Bibhas Chakraborty$^{1,5,6,*}$\email{bibhas.chakraborty@duke-nus.edu.sg}\\
 	$^{1}$ Centre for Quantitative Medicine, Duke-NUS Medical School, Singapore\\
 	$^{2}$ Department of Statistics, University of California, Irvine, Irvine, California, U.S.A.\\
       $^{3}$ Medical Research Council Biostatistics Unit, University of Cambridge, Cambridge, United Kingdom\\
       $^{4}$ Department of Medical Statistics, The London School of Hygiene and Tropical Medicine,\\
       London, United Kingdom \\
 	$^{5}$ Department of Statistics and Data Science, National University of Singapore, Singapore\\
 	$^{6}$ Department of Biostatistics and Bioinformatics, Duke University, Durham, North Carolina, U.S.A.
}

\begin{document}





\pagerange{\pageref{firstpage}--\pageref{lastpage}} 




\label{firstpage}


\begin{abstract}
In mobile health, tailoring interventions for real-time delivery is of paramount importance.  Micro-randomized trials have emerged as the ``gold-standard'' methodology for developing such interventions. Analyzing data from these trials provides insights into the efficacy of interventions and the potential moderation by specific covariates. The ``causal excursion effect", a novel class of causal estimand, addresses these inquiries. Yet, existing research mainly focuses on continuous or binary data, leaving count data largely unexplored. The current work is motivated by the Drink Less micro-randomized trial from the UK, which focuses on a zero-inflated proximal outcome, i.e., the number of screen views in the subsequent hour following the intervention decision point. To be specific, we revisit the concept of causal excursion effect, specifically for zero-inflated count outcomes, and introduce novel estimation approaches that incorporate nonparametric techniques. Bidirectional asymptotics are established for the proposed estimators. Simulation studies are conducted to evaluate the performance of the proposed methods. As an illustration, we also implement these methods to the Drink Less trial data.
\end{abstract}

%



\maketitle


%

\section{INTRODUCTION}
\label{intro}

Mobile health (mHealth) interventions, particularly those utilizing text messages or push notifications, have been developed to promote health-related behaviors. These interventions exhibit significant potential across a wide range of health issues, from alcohol consumption reduction \citep{song2019mobile} to physical activity maintenance \citep{lee2019efficacy}. Advancements in mobile and sensing technologies now facilitate real-time tracking of an individual's internal state and context, offering timely and personalized support \citep{kumar2013mobile, spruijt2014dynamic}. This has given rise to the concept of the just-in-time adaptive intervention (JITAI), which tailors treatment in response to the evolving needs and situations of the individual, bearing the goal of delivering the right treatment on the right occasion \citep{nahum-shani2018}. For instance, evenings are recognized as a high-risk window for individuals with a history of excessive alcohol consumption \citep{day2014gender,bell2020a}. Moreover, the burden on users due to overly frequent notifications can result in disengagement \citep{bell2020a}. Therefore, JITAIs for alcohol use could leverage evenings and the history of notification delivery as tailoring variables to determine the likelihood of delivering an intervention at the current moment.

The micro-randomized trial (MRT) has emerged as the touchstone methodology for devising these interventions  \citep{klasnja2015,liao2016,liu2023microrandomized,qian2022microrandomized}. Within an MRT, participants undergo sequential randomization, aligning them with one of the intervention options across hundreds or even thousands of decision points. Typically, data analysis from micro-randomized trials seeks to answer three critical scientific questions: (1) which interventions can impact the proximal outcome; (2) in which time-varying context should the intervention be delivered; and (3) does the treatment effect change with time? The ``causal excursion effect", a novel class of causal estimand, provides solutions to these inquiries \citep{boruvka2018,qian2021a}. Notably, these effects can be perceived as a marginal generalization of the treatment ``blip" in the \textit{structural nested mean model} \citep{robins1989analysis,robins1994correcting}, as they are conditional on a few selected variables instead of all past observed variables. 

Standard methods like generalized estimating equations (GEE) \citep{liang1986longitudinal} or random-effects models \citep{laird1982random} may not provide consistent treatment effect estimates when data includes time-varying treatments and confounders \citep{sullivan1994cautionary,robins2008estimation}. Structural nested mean models \citep{robins1994correcting} and G-estimation address this by modeling the causal effect of time-varying treatments on time-varying outcomes. \citet{boruvka2018} and \citet{qian2021a} innovatively developed weighted and centered least square estimators for estimating causal excursion effects, but focused on continuous or binary data, leaving count data unexplored. Our study is motivated by the Drink Less trial \citep{bell2020a} where the proximal outcome, the number of screen views following notifications, is zero-inflated count data, invalidating common distributional assumptions.


Building upon \citet{yu2023}, we propose new methods for estimating causal excursion effects under potentially zero-inflated count outcomes, incorporating nonparametric nuisance estimation. Addressing technical errors in collecting randomization probabilities \citep{shi2023metalearning}, we propose doubly robust estimating equations applicable to both MRTs and observational mHealth studies with non-randomized treatments.  Our contributions are two-fold: 1) advocating nonparametric techniques like generalized additive models and two-part models \citep{yu2023} for nuisance estimation under zero-inflation; 2) proposing doubly robust estimators with bidirectional asymptotics \citep{yu2023} as the sample size or the number of decision points increases.


The rest of this article is organized as follows. Section \ref{example} provides an overview of the Drink Less trial as a motivating example. The notation and causal estimand are reviewed in Section \ref{rev}. Section \ref{methods} details the incorporation of nonparametric methods for estimating the causal excursion effect.
Section \ref{simus} provides the data generation procedure, settings, and results of the simulation study. Next, we apply the proposed method to analyzing the Drink Less data in Section \ref{data}. We conclude with a discussion in Section \ref{discuss}. 

\section{MOTIVATING EXAMPLE: DRINK LESS}
\label{example}
Drink Less is a behavior change app that aims to help the general adult population in the UK who want to reduce hazardous and harmful alcohol consumption \citep{garnett2019development,garnett2021refining,bell2020engagement}. A 30-day MRT with 349 participants was conducted to improve the push notification strategy in Drink Less \citep{bell2020a,bell2023}. 


Participants with a baseline Alcohol Use Disorders Identification Test (AUDIT) score of 8 or higher \citep{bohn1995alcohol,saunders1993development}, who resided in the UK, were at least 18 years old, and desired to drink less, were recruited into the trial. Every day at 8 p.m., during the trial, participants were randomly given one of three intervention options: no notification, the standard notification, or a message randomly selected from a new message bank. For more details about the messages, see \citet{bell2020a}. Each option was assigned according to a static randomization probability of 40\%, 30\%, and 30\%, respectively. In this study, we employ the total number of screen views between 8 p.m. and 9 p.m. as the proximal outcome to measure the depth of engagement with the app. Prior research has validated the use of screen views as a metric for near-term engagement \citep{olga2019a,radin2018}. Along with the interventions, data on age, AUDIT score, gender, and other covariates were also recorded.
The main goal is to explore the effects of push notifications on user engagement, whether or not these effects differ according to the user's context, and how these effects change over time. 

\begin{figure}[ht]
	\centering
	\subfigure[Histogram]{\includegraphics[width=7cm,height=5cm,keepaspectratio]{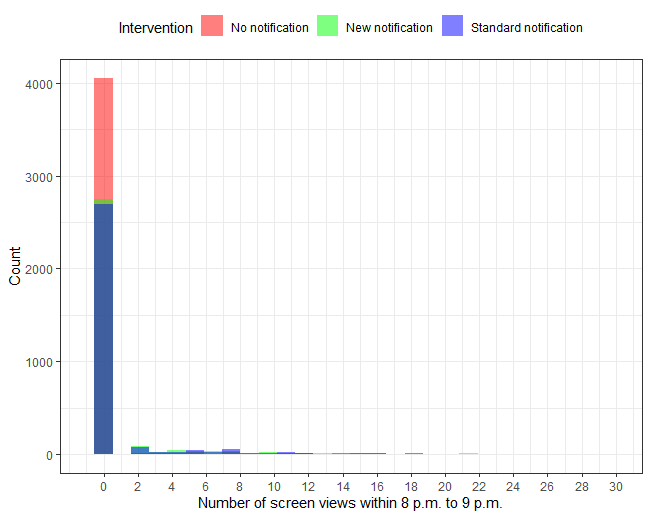}} 
	\subfigure[Q-Q plot]{\includegraphics[width=7cm,height=5cm,keepaspectratio]{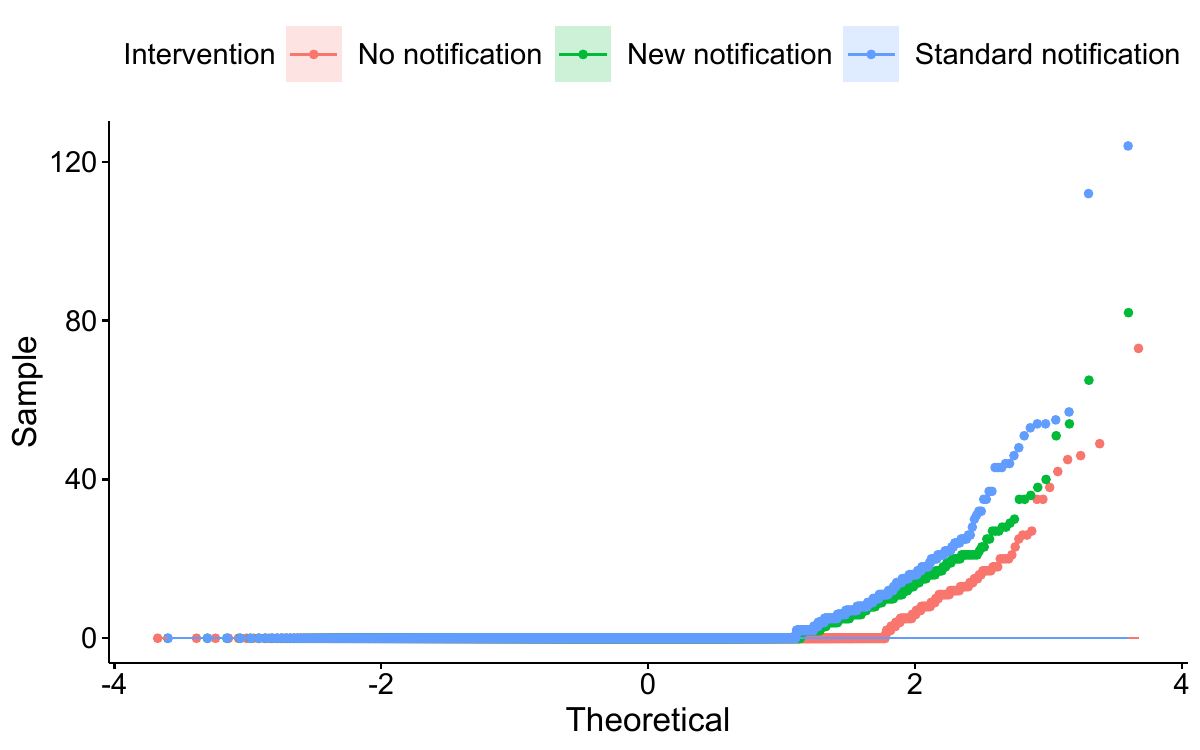}} 
	\caption{Graphical inspection of the distribution and normality check of the number of screen views from 8 p.m. to 9 p.m. after each type of notification.}
	\label{fig:normality}
\end{figure}

We first examine the distribution of the proximal outcome from the Drink Less trial. Figure \ref{fig:normality} displays the histogram and Q-Q plot of the proximal outcome, which is the number of screen views between 8 p.m. and 9 p.m. following no notification, a new notification, or the standard notification. The results suggest that the proximal outcome is not normally distributed and is highly zero-inflated.

\section{CAUSAL EXCURSION EFFECT: A REVIEW}
\label{rev}

\subsection{Notations}
Consider a setting with longitudinal data spanning $T$ decision points for $n$ participants. For each participant, the treatment assignment at time $t$ is represented by $A_t$.  We simplify by considering only binary treatments $A_t \in \{0,1\}$, where $1$ indicates the administration of treatment, and $0$ indicates its absence.  Detailed discussions on extensions to scenarios with multi-category treatments can be found in Appendix \ref{multi}. Let $X_t$ denote individual and contextual information or covariates collected after time $t-1$ and up to time $t$. This includes prior treatments, proximal outcomes, and the participant's availability status $I_t \in \{0,1\}$. The value $1$ implies availability for treatment at $t$, and $0$ denotes the opposite. 

After providing treatment at time $t$, we observe a proximal outcome $Y_{t,\Delta}$. This outcome is a deterministic function of data collected over an interval of length $\Delta$. For example, in the Drink Less study (Section \ref{example}), the decision time is daily at 8 p.m., with the proximal outcome being screen views between 8 p.m. and 9 p.m., resulting in $\Delta = 1$. Other mHealth studies might have larger $\Delta$ values with each minute as a decision point \citep{battalio2021sense2stop}. We focus on situations where $\Delta = 1$ and assume $Y_{t,1}$ is zero-inflated count data.

In our notation, an overbar signifies a sequence of random variables; for instance, $\overline{A}_t$ encompasses the series $(A_1,\cdots,A_t)$. The data collected until time $t$ is represented by the history $H_t = (\overline{X}_t, \overline{A}_{t-1})$. 


\subsection{Causal Excursion Effect}
In the following, we introduce the potential outcomes framework \citep{rubin1974estimating,robins1989analysis} to define the causal excursion effect. Let $X_t(\overline{a}_{t-1})$ be the potential covariates that would have been collected, and $A_t(\overline{a}_{t-1})$ the treatment that would have been assigned, had the participant received the treatment sequence $\overline{a}_{t-1} \in \{0,1\}^{t-1}$. Additionally, denote by $Y_{t,1}(\overline{a}_{t})$ the potential proximal outcome that would have been observed had that participant received the treatment sequence $\overline{a}_{t} \in \{0,1\}^{t}$. Here, treatments $A_t$ and covariates $X_t$ are also expressed as potential outcomes of previous treatment to mimic mHealth settings where covariates and treatment assignments can depend on previous treatments. The potential history at time $t$ is represented by $H_t(\overline{A}_{t-1})=\{X_1,A_1,X_2(A_1),A_2,X_3(\overline{A}_2),\cdots,X_t(\overline{A}_{t-1})\}$. 

As in \citet{boruvka2018} and \citet{qian2021a}, we are interested in estimating the causal excursion effect of treatment $A_t$ on $Y_{t,1}$:
\begin{eqnarray}
    \log \frac{\mathbb{E}\left\{Y_{t, 1}\left(\overline{A}_{t-1}, 1\right) \mid S_t\left(\overline{A}_{t-1}\right) = s_t, I_t\left(\overline{A}_{t-1}\right)=1\right\}}{\mathbb{E}\left\{Y_{t, 1}\left(\overline{A}_{t-1}, 0\right) \mid S_t\left(\overline{A}_{t-1}\right) = s_t, I_t\left(\overline{A}_{t-1}\right)=1\right\}}.
\end{eqnarray}
Here, $S_t\left(\overline{A}_{t-1}\right) \in \mathbb{R}^p$ denotes a vector of potential moderators formed from $H_t(\overline{A}_{t-1})$. To accommodate possibly zero-inflated count outcomes, we opt for the logarithm of the ratio of, rather than the difference in, the expected outcomes. The effect contrasts two excursions from the treatment protocol before time $t$ and characterizes the treatment effect in the short term. Specifically, it assesses the effect of following the protocol until time $t-1$ and then deviating at time $t$ to assign treatment $1$, compared to a deviation that assigns treatment $0$ at the same point. By conditioning on $I_t\left(\overline{A}_{t-1}\right)=1$ and $S_t\left(\overline{A}_{t-1}\right)$, the effect is defined for only individuals available for treatment at time $t$ and is marginalized over variables in $H_t(\overline{A}_{t-1})$ that are not in $S_t(\overline{A}_{t-1})$. It is critical to recognize that the causal excursion effect depends on the treatment assignment protocol $\overline{A}_{t-1}$, diverging from traditional causal inference literature \citep{robins1994correcting}. This feature is particularly valuable in the context of real-life trials, where the impact of past treatments on user burden or disengagement can lead to significantly varied treatment effects at time $t$ \citep{qian2021c}. Consequently, the causal excursion effect provides a framework for improving the current treatment protocol. Setting $S_t(\overline{A}_{t-1})$ to an empty set allows for the assessment of an intervention's direct effect on proximal outcomes. Incorporating moderators such as days since download, treatment history, or previous outcomes further enables the evaluation of treatment effect moderation. These insights are pivotal for future mobile health application development and treatment policy formulation \citep{Luckett2020}.

When choosing $S_t\left(\overline{A}_{t-1}\right) = H_t\left(\overline{A}_{t-1}\right)$, we get the fully conditional version of the causal excursion effect, which can be expressed as
\begin{eqnarray}
    \log \frac{\mathbb{E}\left\{Y_{t, 1}\left(\overline{A}_{t-1}, 1\right) \mid H_t\left(\overline{A}_{t-1}\right), I_t\left(\overline{A}_{t-1}\right)=1\right\}}{\mathbb{E}\left\{Y_{t, 1}\left(\overline{A}_{t-1}, 0\right) \mid H_t\left(\overline{A}_{t-1}\right), I_t\left(\overline{A}_{t-1}\right)=1\right\}}.
\end{eqnarray}
The fully conditional effect closely parallels the treatment blips in the structural nested mean model \citep{robins1994correcting}. However, our focus is solely on the immediate effect of a time-varying treatment, not the cumulative effect of all previous treatments. 

\subsection{Identification}
To estimate the causal excursion effect from the observed data, we state three fundamental assumptions in causal inference.
\begin{assumption}[Consistency]
\label{ass1}
The observed data is equal to the potential outcome under the observed treatment sequence, i.e., for $2 \leq t \leq T$, $X_t = X_t(\overline{A}_{t-1})$, $A_t=A_t(\overline{A}_{t-1})$, $Y_{t,1}=Y_{t,1}(\overline{A}_{t})$. 
\end{assumption}
\begin{assumption}[Positivity]
\label{ass2}
If the joint density $\operatorname{Pr}(H_t=h_t,I_t=1)>0$, then $\operatorname{Pr}(A_t=a_t|H_t=h_t,I_t=1)>0$ almost everywhere for $a_t \in \{0,1\}$.
\end{assumption}
\begin{assumption}[Sequential ignorability]
\label{ass3}
For $1\leq t \leq T$, the potential outcomes $\{X_{t+1}(\overline{a}_t),\allowbreak A_{t+1}(\overline{a}_t),  \allowbreak  \ldots, X_{T+ 1}(\overline{a}_T\allowbreak )\}$ are independent of $A_t$ conditional on $H_t$.
\end{assumption}

Assumption \ref{ass1} connects the potential outcomes with the observed data and states that there is no interference between the observations. Both Assumptions \ref{ass2} and \ref{ass3} are inherently satisfied in MRTs due to the sequential randomization of treatments based on known probabilities. 

Under these assumptions, 
the causal excursion effect can be expressed in terms of observed data:
\begin{eqnarray}
    \log \frac{\mathbb{E}\left[\mathbb{E}\left\{ Y_{t, 1} \mid A_t=1, H_t, I_t=1\right\} \mid S_t, I_t=1\right]}{\mathbb{E}\left[\mathbb{E}\left\{ Y_{t, 1} \mid A_t=0, H_t, I_t=1\right\} \mid S_t, I_t=1\right]}.
\end{eqnarray}
The derivation of the identifiability results follows similarly to \citet{qian2021a}. For completeness, the proof for the general case where $\Delta > 1$ is included in Appendix \ref{sup1}.

Additionally, the causal excursion effect simplifies to 
\begin{eqnarray}
    \log \frac{\mathbb{E}\left(Y_{t, 1} \mid A_t=1, H_t, I_t=1\right)}{\mathbb{E}\left(Y_{t, 1} \mid A_t=0, H_t, I_t=1\right)} .
\end{eqnarray}
when conditioning on the full history $H_t$. In what follows, we begin by detailing the estimation procedure for the fully conditional excursion effect, subsequently extending the approach to instances where $S_t\neq H_t$. 

\section{INCORPORATING NONPARAMETRIC METHODS FOR CAUSAL EXCURSION EFFECT ESTIMATION}
\label{methods}
\subsection{Estimating the Conditional Excursion Effect} 
In this section, we discuss the estimation procedure for the fully conditional excursion effect, drawing inspiration from the G-estimator initially introduced by \citet{robins1994correcting} and subsequently discussed by \citet{yu2023} for the multiplicative structural nested mean model. Our approach differs in several aspects: it centers on estimating the causal excursion effect, which serves as a marginal extension of the treatment ``blip'' in structural nested mean models, as detailed in Section \ref{est_mar}. Additionally, it introduces a weighting component to enable semi-parametric efficiency or facilitate the estimation of marginal treatment effects. The addition of an availability indicator, $I_t$, further adapts our model for effective application in mHealth contexts. 

We begin by assuming a parametric model for the conditional excursion effect:
\begin{eqnarray}
\label{mod1}
    \log \frac{\mathbb{E}\left(Y_{t, 1} \mid A_t=1, H_t, I_t=1\right)}{\mathbb{E}\left(Y_{t, 1} \mid A_t=0, H_t, I_t=1\right)}  = f(H_t)^{\top}\phi
\end{eqnarray}
for unknown $p$-dimensional parameter vector $\phi$, where $f(\cdot)$ is a known deterministic function. Throughout, we denote the true value of $\phi$ by $\phi_0$. 

Then, define $U_t(\phi) = Y_{t,1} \exp(-A_tf(H_t)^{\top}\phi)$ to mimic the potential outcome $Y_{t,1}(\overline{A}_{t-1},0)$ that would have been observed had the treatment been removed at time $t$. 
By Assumption \ref{ass3},
\begin{eqnarray}
    \mathbb{E}[U_t(\phi)|A_t, H_t, I_t=1] &=& \mathbb{E}[Y_{t,1}(\overline{A}_{t-1},0)|A_t,H_t, I_t=1]\nonumber \\
    &=& \mathbb{E}[Y_{t,1}(\overline{A}_{t-1},0)|H_t, I_t=1] \nonumber \\
    &=& \mathbb{E}[U_t(\phi)|H_t, I_t=1].  \label{lem}
\end{eqnarray}
Let $h_0(H_t) = \mathbb{E}[U_t(\phi)|H_t, I_t=1]$ denote the conditional mean of $U_t(\phi)$ given $H_t$ and $I_t=1$, and let $p_{t,0}(H_t)=\operatorname{Pr}(A_t=1|H_t)$ denote the treatment randomization probability. 
Estimation of $\phi$ in model \eqref{mod1} can thus be based on the following estimating function:
\begin{eqnarray}
\label{est0}
    m_C(H_{t};\phi, h, p_t) = I_t \left[U_t(\phi) - h(H_t)\right]\times \left[\{A_t - p_t(H_t)\}f(H_t)\right],
\end{eqnarray}
for any given  $h$ and $p_t$. Building upon argument \eqref{lem}, the expectation of \eqref{est0} equals zero when $\phi=\phi_0$ under conditions $h = h_0$ or $p_t = p_{t,0}$. To denote individual participants, we introduce the subscript $i$, resulting in $n$ independent and identically distributed copies of the data sequence $H_T$, represented as $H_{1,T}, \ldots, H_{n,T}$. The estimating equation for $\phi$ is expressed as:
\begin{eqnarray}
    \frac{1}{nT}\sum_{i=1}^n\sum_{t=1}^T m_C(H_{it};\phi, h, p_t) = 0.
\end{eqnarray}
In MRTs with known randomization probabilities, estimating the conditional mean $h_0(H_t)$ is the primary task. We express it as:
\begin{eqnarray*}
    h_0(H_t) = \mu_{1,0}(H_t)\exp\{-f(H_t)^{\top}\phi\}p_{t}(H_t) +  \mu_{0,0}(H_t)\left(1-p_{t}(H_t)\right),
\end{eqnarray*}
where $\mu_{1,0}(H_t) = \mathbb{E}[Y_{t,1}|A_t=1,H_t, I_t=1]$ and $\mu_{0,0}(H_t) = \mathbb{E}[Y_{t,1}|A_t=0,H_t, I_t=1]$. Nonparametric models such as generalized additive models can be employed to estimate $\mu_{1,0}(H_t)$ and $\mu_{0,0}(H_t)$. Plugging estimates $\widehat{\mu}_1$ and $\widehat{\mu}_0$ into \eqref{est0} yields the estimating function $m_C(H_t; \phi, \widehat{\mu}_1, \widehat{\mu}_0, p_{t,0})$. Denote $\eta=(\mu_0, \mu_1, p_t)$, the true value $\eta_0 = (\mu{0,0}, \mu{1,0}, p{t,0})$, and its estimator $\widehat{\eta}=(\widehat{\mu}_0, \widehat{\mu}_1, \widehat{p}_t)$.  For simplicity, denote $m_C(H_t; \phi, \eta) = m_C(H_t; \phi, \mu_1, \mu_0, p_t)$.  This resulting estimator is referenced as ``ECE-NonP'', following the terminology used by \citet{qian2021a}.

In observational studies, both $h_0(H_t)$ and $p_{t,0}(H_t)$ are unknown, necessitating estimation of these nuisance functions from data. Since we cannot guarantee consistent nuisance estimators, a doubly robust estimator ensuring consistency when either $h_0(H_t)$ or $p_{t,0}(H_t)$ is consistently estimated becomes crucial. We demonstrate the rate double robustness of our proposed ECE-NonP estimator below.

Following \citet{yu2023}, we make an additional assumption:
\begin{assumption}
    \label{ass4}
    As $n T \rightarrow \infty,\left\|\widehat{\eta}-\eta_0\right\|_{2, P} \rightarrow 0$ in probability and
\begin{eqnarray*}
\left(\left\|\widehat{\mu}_1-\mu_{1,0}\right\|_{2, P}+\left\|\widehat{\mu}_0-\mu_{0,0}\right\|_{2, P}+\left\|\widehat{p}_t-p_{t,0}\right\|_{2, P}\right)\left\|\widehat{p}_t-                                               p_{t,0}\right\|_{2, P}=o_P\left\{(n T)^{-1 / 2}\right\} .
\end{eqnarray*}
\end{assumption}
which outlines nuisance function convergence rates for establishing asymptotics. The first part ensures estimator consistency, while the second part determines the ECE-NonP asymptotic distribution. For MRT data, this assumption is naturally met as $p_{t,0}(H_t)$ is known, and others can be consistently estimated using nonparametric methods. Theorem \ref{th1} provides bidirectional asymptotics for ECE-NonP, with the proof presented in Appendix \ref{sup2}.

\begin{theorem}[Bidirectional Asymptotics of ECE-NonP]
\label{th1}
    Suppose model \eqref{mod1}, Assumptions \ref{ass1} - \ref{ass3}, Assumption \ref{ass4} and some regularity conditions hold. Let ${\dot m}_\phi(H_t; \phi, \eta) = \frac{\partial m_C(H_t; \phi,\eta)}{\partial \phi}$. Let $\mathbb{P}g\{m_C(H_{t}; \phi, \eta)\} = \frac{1}{T}\sum_{t=1}^{T} \mathbb{E}[g\{m_C(H_{i,t};\phi, \eta)\}]$ where $g$ is any given function or operator of $m_{C}$. 
    As either $n \rightarrow \infty$ or $T \rightarrow \infty$, we have $(nT)^{1/2}(\widehat{\phi} - \phi_0) \rightarrow \operatorname{MVN}(0, \bmath{B} \bmath{\Sigma} \bmath{B}^{\top})$ in distribution, where $\bmath{B} = \{\mathbb{P}{\dot m}_\phi(H_{t};\phi_0,\eta_0)\}^{-1}$ and
    \begin{eqnarray*}
        \bmath{\Sigma} = \lim _{n T \rightarrow \infty} \frac{1}{n T} \sum_{i=1}^n \sum_{t=1}^{T} \mathbb{E}\left\{m_{C}\left(H_{i,t};\phi_0, \eta_0\right) m_{C}^{\top}\left(H_{i,t}; \phi_0, \eta_0\right)|H_{i,t}\right\}.
    \end{eqnarray*}
\end{theorem}

\begin{remark}
    \citet{qian2021a} introduced an estimator ECE for the conditional excursion effect with binary outcomes. Following \citet{kim2021a}, ECE can be extended to count outcomes using the estimating function: 
    \begin{eqnarray*}
        &&I_t \exp \left\{-A_t f\left(H_t\right)^{\top} \phi\right\}\left[Y_{t, 1}-\exp \left\{g\left(H_t\right)^{\top} \alpha+A_t f\left(H_t\right)^{\top} \phi\right\}\right]\\
       && \times \widetilde{K}_t\left[\begin{array}{c}
g\left(H_t\right) \\
\left\{A_t-p_t\left(H_t\right)\right\} f\left(H_t\right)
\end{array}\right],
    \end{eqnarray*}
    where $\widetilde{K}_t$ is defined as
    \begin{eqnarray*}
        -\frac{\exp \left\{f\left(H_t\right)^{\mathrm{T}} \phi\right\}}{\exp \left\{f\left(H_t\right)^{\mathrm{T}} \phi\right\} p_t\left(H_t\right)+\left\{1-p_t\left(H_t\right)\right\}}.
    \end{eqnarray*}
\end{remark}
\noindent Their estimator assumes a parametric working model for $\mathbb{E}(Y_{t,1}|H_t, A_t=0, I_t=1)$. Instead, we represent $h_0(H_t)$ using nonparametrically estimated $\widehat{\mu}_{1}(H_t)$ and $\widehat{\mu}_{0}(H_t)$. Semiparametric efficiency is achieved when the working model is correctly specified. We examine ECE and ECE-NonP's performance with the above weight $\widetilde{K}_t$ in simulations.

\begin{remark}
    The ECE-NonP estimator's consistency relies on correctly specifying the treatment effect model \eqref{mod1}.  However, the high dimensionality of the full history $H_t$ complicates accurate specification of the model. Currently, there is limited guidance on how to accurately determine the treatment effect model, which remains a prominent challenge.
\end{remark}

\begin{remark}
    The principle of bidirectional asymptotics suggests that as either the sample size $n$ or the number of decision points $T$ approaches infinity, the ECE-NonP estimator achieves consistency and asymptotic normality. However, this should be used with caution. A larger $T$ may not necessarily imply more information for estimating $\phi$ when considering user availability during a trial. Time points where $I_t=0$ do not contribute to \eqref{est0}, and the value of $I_t$ can vary over time. This is not a problem in studies like Drink Less, where users are assumed to be always available given the intervention nature.
\end{remark}

\subsection{Estimating the Marginal Excursion Effect}
\label{est_mar}
For the marginal excursion effect, we assume the following parametric model for $1 \leq t \leq T$:
\begin{eqnarray}
\label{mod2}
        \log \frac{\mathbb{E}\left[\mathbb{E}\left\{ Y_{t, 1} \mid A_t=1, H_t, I_t=1\right\} \mid S_t, I_t=1\right]}{\mathbb{E}\left[\mathbb{E}\left\{ Y_{t, 1} \mid A_t=0, H_t, I_t=1\right\} \mid S_t, I_t=1\right]} = S_t^{\top}\beta
\end{eqnarray}
where $\beta$ is the parameter of interest with true value $\beta_0$. A linear model is assumed for ease of interpretation, but the procedure extends to nonlinear forms \citep{qian2021a}.

Then, we construct a variable $U_t = Y_{t,1}\exp\left(-A_tS_t^{\top}\beta\right)$ under model \eqref{mod2}. Estimation of $\beta$ can be based on the following estimating function:
\begin{eqnarray}
\label{est1}
    m_M(H_t; \beta, \eta) = I_t W_t\left[U_t(\beta) - h(H_t)\right]\times \left[\{A_t - \widetilde{p}_t(S_t)\}S_t\right],
\end{eqnarray}
where weight $W_t$ is added to allow for the estimation of marginal causal excursion effects conditional on $S_t$ instead of $H_t$. The weight $W_t$ is defined as 
\begin{eqnarray*}
    \left\{\frac{\widetilde{p}_t\left(S_t\right)}{p_t\left(H_t\right)}\right\}^{A_t}\left\{\frac{1-\widetilde{p}_t\left(S_t\right)}{1-p_t\left(H_t\right)}\right\}^{1-A_t} 
\end{eqnarray*}
where the numerical probability $\widetilde{p}_t\left(S_t\right)$ can be chosen arbitrarily as long as it only depends on $S_t$. Intuitively, the weight $W_t$ transforms the data distribution where $A_t$ is randomized with probability $p_t(H_t)$ to a distribution where $A_t$ is randomized with probability $\widetilde{p}_t(S_t)$. 
The solution to $\frac{1}{nT}\sum_{i=1}^n\sum_{t=1}^{T}m_M(H_{i,t}; \beta, \eta) = 0$ yields an estimator for $\beta$. 

Similar to before, we need to estimate the nuisance function $h_0(H_t)$ and plug estimates into \eqref{est1}.  We can express $h_0(H_t)$ using two conditional means:
\begin{eqnarray*}
    h_0(H_t) = \mu_{1,0}(H_t) \exp\{-S_t^{\top}\beta\}\widetilde{p}_t(S_t) +  \mu_{0,0}(H_t)\left(1-\widetilde{p}_t(S_t)\right),
\end{eqnarray*}
where $p_t(H_t)$ is replaced by $\widetilde{p}_t(S_t)$ due to the change of probability induced by weight $W_t$.  In MRTs, we assume the propensity score $p_t(H_t)$ appearing in $W_t$ is known. We label this the ``EMEE-NonP" approach.

Theorem \ref{prop2} provides the bidirectional asymptotics of the EMEE-NonP estimator, proven in Appendix \ref{sup3}. 

\begin{theorem}[Bidirectional Asymptotics of EMEE-NonP]
\label{prop2}
    Suppose the randomization probability $p_t(H_t)$ is known. Suppose model \eqref{mod2}, Assumptions \ref{ass1} - \ref{ass3}, Assumption \ref{ass4} and some regularity conditions hold. Let ${\dot m}_\beta(H_t; \beta, \eta) = \frac{\partial m_M(H_t; \beta,\eta)}{\partial \beta}$. Let $\mathbb{P}g\{m_M(H_{t}; \beta, \eta)\} = \frac{1}{T}\sum_{t=1}^{T}\allowbreak \mathbb{E}[g\{m_M(H_{i,t};\beta, \eta)\}]$ where $g$ is any given function or operator of $m_{M}$. 
    As either $n \rightarrow \infty$ or $T \rightarrow \infty$, we have $(nT)^{1/2}(\widehat{\beta} - \beta_0) \rightarrow \operatorname{MVN}(0, \bmath{B} \bmath{\Sigma} \bmath{B}^{\top})$ in distribution, where $\bmath{B} = \{\mathbb{P}{\dot m}_\beta(H_{t};\beta_0,\eta_0)\}^{-1}$ and
    \begin{eqnarray*}
        \bmath{\Sigma} = \lim _{n T \rightarrow \infty} \frac{1}{n T} \sum_{i=1}^n \sum_{t=1}^{T} \mathbb{E}\left\{m_{M}\left(H_{i,t};\beta_0, \eta_0\right) m_{M}^{\top}\left(H_{i,t}; \beta_0, \eta_0\right) \mid H_{i,t}\right\}.
    \end{eqnarray*}
\end{theorem}

\begin{remark}
The EMEE method, as proposed in \citet{qian2021a} for a binary outcome, can concurrently determine both $\alpha$ and $\beta$. The estimating equation is expressed as:
\begin{eqnarray}
\label{est4}
    \sum_{t=1}^{T} I_t \exp \left\{-A_t S_t^{\top} \beta\right\}\left[Y_{t, 1}-\exp \left\{g\left(H_t\right)^{\top} \alpha+A_t S_t^{\top} \beta\right\}\right] W_t\left[\begin{array}{c}
g\left(H_t\right) \\
\left\{A_t-\widetilde{p}_t\left(S_t\right)\right\} S_t
\end{array}\right]
\end{eqnarray}
One important property of the estimator is that the consistency is robust to the misspecification of the working model $g(H_t)^{\top}\alpha$. In our method, the essential requirement for robustness is embodied in Assumption \ref{ass4}.
\end{remark}

\begin{remark}
The EMEE-NonP estimator lacks the double robustness property, which we demonstrate:
    \begin{eqnarray*}
        &&\mathbb{E}\left[S_tI_tW_t\left\{U_t(\beta) - h(H_t)\right\}\times \left\{A_t-\widetilde{p}_t(S_t)\right\}\right]\\
        &=& \mathbb{E}\left[\mathbb{E}\left\{S_tI_tW_t\left(U_t(\beta) - h(H_t)\right)\times \left(A_t-\widetilde{p}_t(S_t)\right)|A_t, H_t\right\}\right]\\
        &=& \mathbb{E}\left\{ S_tI_t\mathbb{E}\left[W_t |A_t, H_t\right] \times  \left[\mathbb{E}(U_t(\beta)|A_t, H_t, I_t=1) - h(H_t)\right] \times \left[p_t(H_t)-\widetilde{p}_t(S_t)\right]\right\}
    \end{eqnarray*}
    For double robustness, the condition $\mathbb{E}(U_t(\beta)|A_t, H_t,I_t=1) = h_0(H_t) = \mathbb{E}[U_t(\beta)|H_t, I_t=1]$ is required. However, under Assumptions \ref{ass1} - \ref{ass3}, this cannot be ensured. A similar issue arises for the EMEE estimator. As highlighted in \citet{qian2021c}, even if $\exp \left\{g\left(H_t\right)^{\top} \alpha\right\}$ correctly models $\mathbb{E}\left\{Y_{t, 1}\left(\overline{A}_{t-1}, 0\right) \mid H_t, I_t=1, A_t=0\right\}$, the term $Y_{t, 1}-\exp \left\{g\left(H_t\right)^{\top} \alpha+A_t S_t^{\top} \beta\right\}$ in the estimating equation \eqref{est4} does not generally have conditional expectation zero given $H_t$. This occurs because while the causal excursion effect exists as a marginal model, the fully conditional effect, dependent on $H_t$, is not necessarily $S_t^{\top} \beta$. 
\end{remark}

\begin{remark}
For the estimator of $\beta$ to be consistent, the treatment effect model defined in \eqref{mod2} must be correctly specified. Since $S_t$ consists of summary variables chosen from $H_t$ by the researcher, this is an easier task compared to modeling the conditional excursion effect. Yet, by conditioning on only part of the history, i.e., $S_t$, the marginal excursion effect depends on the randomization probability of past treatment assignments $\overline{A}_{t-1}$ \citep{guo2021discussion,zhang2021a,qian2021c}. Therefore, any interpretation of the causal excursion effect must be contextualized within the current treatment protocol.
\end{remark}

\subsection{A Doubly-Robust EMEE-NonP Estimator}
The preceding EMEE-NonP estimator lacks double robustness, which means that it solely relies on known or correctly specified randomization probabilities. This makes it difficult to extend to observational mHealth studies. In this section, we provide a doubly robust estimator, referred to as the DR-EMEE-NonP estimator. 

The modified estimating function is expressed as
\begin{eqnarray}
    m_D(H_t; \beta, \eta) &=& I_t W_t\left\{U_t(\beta) - \mu_{A_t}(H_t)\exp(-A_tS_t^\top\beta)\right\}\times [\{A_t - \widetilde{p}_t(S_t)\}S_t] \nonumber \\
    &+& I_t\widetilde{p}_t(S_t)(1-\widetilde{p}_t(S_t))\left\{\mu_{1}(H_t)\exp(-S_t^{\top}\beta)-\mu_0(H_t)\right\}S_t,
\end{eqnarray}
where $\mu_{A_t}(H_t) = \mathbb{E}[Y_{t,1}|H_t,A_t,I_t=1]$.

Below we make a new assumption about the convergence rate of nuisance functions and establish the bidirectional asymptotics for DR-EMEE-NonP in Theorem \ref{prop3}. The proof is deferred to Appendix \ref{sup4}. 

\begin{assumption}
\label{ass5}
As $n T \rightarrow \infty,\left\|\widehat{\eta}-\eta_0\right\|_{2, P} \rightarrow 0$ in probability and
   \begin{eqnarray*}
        \left(\|\widehat{\mu}_1 - \mu_{1,0}\|_{2,P} + \|\widehat{\mu}_0 - \mu_{0,0}\|_{2,P}\right)\|  \widehat{p}_t - p_{t,0}\|_{2,P} = o_P\{(nT)\}^{-1/2}
   \end{eqnarray*}
\end{assumption}

\begin{theorem}[Bidirectional Asymptotics of DR-EMEE-NonP]
\label{prop3}
Suppose model \eqref{mod2}, Assumptions \ref{ass1} - \ref{ass3}, Assumption \ref{ass5} and some regularity conditions hold. Let ${\dot m}_\beta(H_t; \beta, \eta) = \frac{\partial m_D(H_t; \beta,\eta)}{\partial \beta}$. Let $\mathbb{P}g\{m_D(H_{t}; \beta, \eta)\} = \frac{1}{T}\sum_{t=1}^{T} \mathbb{E}[g\{m_D(H_{i,t};\beta, \eta)\}]$ where $g$ is any given function or operator of $m_{D}$. As either $n \rightarrow \infty$ or $T \rightarrow \infty$, we have $(nT)^{1/2}(\widehat{\beta} - \beta_0) \rightarrow \operatorname{MVN}(0, \bmath{B} \bmath{\Sigma} \bmath{B}^{\top})$ in distribution, where $\bmath{B} = \{\mathbb{P}{\dot m}_\beta(H_{t};\beta_0,\eta_0)\}^{-1}$ and
    \begin{eqnarray*}
        \bmath{\Sigma} = \lim _{n T \rightarrow \infty} \frac{1}{n T} \sum_{i=1}^n \sum_{t=1}^{T} \mathbb{E}\left\{m_{D}\left(H_{i,t};\beta_0, \eta_0\right) m_{D}^{\top}\left(H_{i,t}; \beta_0, \eta_0\right) \mid H_{i,t}\right\}.
    \end{eqnarray*}
\end{theorem}
The idea of forming a doubly robust estimator originated with the research of \citet{scharfstein1999adjusting}. This foundational work was subsequently expanded upon by multiple studies, notably the DR-Learner as explored by \citet{van2006targeted}, \citet{nie2021quasi} and \citet{kennedy2023optimal}. More recently, \citet{shi2023metalearning} introduced a DR-learner designed to estimate causal excursion effects in mHealth studies, specifically for continuous and binary outcomes. Our work distinguishes itself in two primary ways. Firstly, the proposed approaches center on estimating the causal excursion effect for zero-inflated count outcomes—an issue previously unexamined. Secondly, we establish bidirectional asymptotics for the DR-EMEE-NonP estimator, necessitating that either the sample size or the number of decision points approaches infinity.

\section{SIMULATION STUDIES}
\label{simus}
In this section, we conduct extensive simulation experiments to assess the finite-sample performance of the proposed estimators across several scenarios:
\begin{enumerate}
    \item[(1)] A binary treatment MRT with $p_t\left( H_t\right)=\operatorname{expit}\left(-0.5 A_{t-1}+ 0.5 Z_t\right)$, where $Z_t$ is a time-varying covariate.
    \item[(2)] A binary treatment observational study with $p_t\left( H_t\right)=\operatorname{expit}\left(-0.5 A_{t-1}+ 0.5 Z_t\right)$.
    \item[(3)] A binary treatment MRT with $p_t\left( H_t\right)$ given by a Thompson sampling (TS) algorithm \citep{russo2018}
    \item[(4)] A three-category treatment MRT with $p_{t,1}\left( H_t\right)=p_{t,2}\left( H_t\right) =0.5\operatorname{expit}\left(-0.5 A_{t-1}+ 0.5 Z_t\right)$
\end{enumerate}
In Scenarios (1)-(2), we compare eight estimators: ECE, ECE-NonP, EMEE, EMEE-NonP, DR-EMEE-NonP, GEE with independence (GEE.IND) or exchangeable (GEE.EXCH) working correlation, and the G-estimator \citep{yu2023} which mirrors ECE-NonP without $\widetilde{K}_t$ and EMEE-NonP without $W_t$. Other scenarios focus on EMEE, EMEE-NonP, DR-EMEE-NonP, and GEE methods.


We adopt a simple setting with $\Delta = 1$, and all participants are available at all decision points, i.e., $I_t=1$, $t=1,\ldots,T$. The time-varying covariate $Z_t$ can take three values, $0$, $1$, and $2$, each with an equal probability. We generate the outcome $Y_{t,1}$ using a zero-inflated negative binomial (NB) model as follows:
\begin{eqnarray*}
    O_{t} &\sim& \operatorname{Bernoulli}(\pi_{t}),\\
    L_{t} &\sim& \operatorname{NB}(\mu_{t}, r),\\
    Y_{t,1} &=& O_{t}L_{t},
\end{eqnarray*}
where $r=1$ denotes the dispersion parameter. 

In Scenario (1), we set $\pi_{t} = \exp(-0.4(Z_t + 0.1) + 0.1 Z_t A_t)$ and $\mu_{t} = \{2.2 \mathbbm{1}_{Z_t=0}+2.5 \mathbbm{1}_{Z_t=1} +2.4 \mathbbm{1}_{Z_t=2}\} \exp\{A_t\left(0.1+0.3 Z_t\right)\}$. Hence, the true conditional causal excursion effect is
\begin{eqnarray*}
    \log \frac{\mathbb{E}\left(Y_{t,1} \mid A_t=1, H_t\right)}{\mathbb{E}\left(Y_{t,1} \mid A_t=0, H_t\right)}=0.1+0.4 Z_t.
\end{eqnarray*}
Here, $Z_t$ interacts with the treatment $A_t$, a moderator that impacts the conditional causal excursion effect. We also consider the fully marginal excursion effect, which is
\begin{eqnarray*}
    \log \frac{\mathbb{E}\left\{\mathbb{E}\left(Y_{t,1} \mid A_t=1, H_t\right)\right\}}{\mathbb{E}\left\{\mathbb{E}\left(Y_{t,1} \mid A_t=0, H_t\right)\right\}}=0.460.
\end{eqnarray*}

For Scenario (2), we set $\pi_{t} = \exp(-0.4(Z_t + 0.1) + 0.1 Z_t A_t)$ and $\mu_{t} = \exp\{0.2 + 0.5Z_t + A_t\left(0.1+0.3 Z_t\right)\}$. In this case, the true marginal excursion effect becomes
\begin{eqnarray*}
    \log \frac{\mathbb{E}\left\{\mathbb{E}\left(Y_{t,1} \mid A_t=1, H_t\right)\right\}}{\mathbb{E}\left\{\mathbb{E}\left(Y_{t,1} \mid A_t=0, H_t\right)\right\}}=0.578.
\end{eqnarray*}
We omit the details of Scenarios (3) and (4) here and defer these to Appendix \ref{sup6}.

The numerator of the weight $W_t$, or the numerical probability $\widetilde{p}_t(S_t)$, is a constant in $t$, given by $\frac{\sum_{n=1}\sum_{t=1}^TA_{i,t}}{Tn}$. A working model $g(H_t)^{\top}\alpha = \alpha_0 + \alpha_1 Z_t$ is used for the logarithm of the expected outcome under no treatment when implementing ECE, EMEE, and GEE methods. For ECE-NonP, EMEE-NonP, and DR-EMEE-NonP, the preliminary step involves estimating the nuisance functions $\mathbb{E}[Y_{t,1}|A_t=1,H_t]$ and $\mathbb{E}[Y_{t,1}|A_t=0,H_t]$ via nonparametric regressions, specifically using generalized additive models. Given the zero-inflation in the data, a hurdle model is used to model the conditional mean in two parts: the probability of attaining value $0$ and the non-zero counts. \citep{hu2011zero}. In the observational data framework, the objective is to examine the rate double robustness of the proposed DR-EMEE-NonP estimator. Here, the randomization probability $p_{t,0}(H_t)$ also requires estimation from the data, for which we use the sample proportion. 

The performance measures include estimation bias (Bias), mean estimated standard error (SE), standard deviation (SD), root mean squared error (RMSE), and coverage probability of 95\% confidence interval (CP) across $1000$ replicates. In the simulation experiments, we set the number of decision points to $T= 30, 100, 150$ and the sample size to $n = 100$.

Table \ref{simu1} presents simulation results of marginal excursion effects under Scenario (1). DR-EMEE-NonP, EMEE-NonP, and EMEE consistently display negligible bias across all settings, with empirical coverage probabilities of 95\% confidence intervals closely aligned to the nominal level. In contrast, ECE-NonP, ECE, and the G-estimator underperform compared to DR-EMEE-NonP, EMEE-NonP, and EMEE, exhibiting greater biases and empirical coverage probabilities of 95\% confidence intervals significantly deviating from the nominal level due to misspecification of the treatment effect model.

\begin{table}
\caption{Comparison of eight estimators for the fully marginal excursion effect under Scenario (1)}
\centering
\resizebox{0.8\textwidth}{!}{%
\begin{tabular}[t]{ccccccc}
\toprule
Estimator & Time Length & Bias & SE & SD & RMSE & CP\\
\midrule
 & 30 & \textbf{-0.025} & 0.055 & 0.056 & 0.061 & \textbf{0.91}\\

 & 100 & \textbf{-0.025} & 0.030 & 0.031 & 0.040 & \textbf{0.86}\\

\multirow{-3}{*}{ ECE} & 150 & \textbf{-0.024} & 0.025 & 0.025 & 0.035 & \textbf{0.83}\\
\cmidrule{1-7}
 & 30 & \textbf{-0.025} & 0.056 & 0.056 & 0.061 & \textbf{0.92}\\

 & 100 & \textbf{-0.025} & 0.030 & 0.031 & 0.040 & \textbf{0.87}\\

\multirow{-3}{*}{ ECE-NonP} & 150 & \textbf{-0.024} & 0.025 & 0.025 & 0.035 & \textbf{0.84}\\
\cmidrule{1-7}
 & 30 & \textbf{-0.013} & 0.057 & 0.058 & 0.059 & 0.93\\

 & 100 & \textbf{-0.013} & 0.032 & 0.032 & 0.035 & \textbf{0.92}\\

\multirow{-3}{*}{G-estimator} & 150 & \textbf{-0.012} & 0.026 & 0.026 & 0.029 & \textbf{0.92}\\
\cmidrule{1-7}
 & 30 & -0.001 & 0.058 & 0.059 & 0.059 & 0.94\\

 & 100 & -0.001 & 0.032 & 0.033 & 0.033 & 0.94\\

\multirow{-3}{*}{ EMEE} & 150 & 0.000 & 0.026 & 0.026 & 0.026 & 0.93\\
\cmidrule{1-7}
 & 30 & -0.001 & 0.058 & 0.058 & 0.058 & 0.95\\

 & 100 & -0.001 & 0.032 & 0.033 & 0.033 & 0.94\\

\multirow{-3}{*}{ EMEE-NonP} & 150 & 0.000 & 0.026 & 0.026 & 0.026 & 0.94\\
\cmidrule{1-7}
 & 30 & -0.001 & 0.058 & 0.059 & 0.058 & 0.95\\

 & 100 & -0.001 & 0.032 & 0.033 & 0.033 & 0.94\\

\multirow{-3}{*}{ DR-EMEE-NonP} & 150 & 0.000 & 0.026 & 0.026 & 0.026 & 0.94\\
\cmidrule{1-7}
 & 30 & \textbf{-0.026} & 0.055 & 0.055 & 0.061 & \textbf{0.91}\\

 & 100 & \textbf{-0.025} & 0.030 & 0.031 & 0.040 & \textbf{0.87}\\

\multirow{-3}{*}{ GEE (ind)} & 150 & \textbf{-0.025} & 0.024 & 0.025 & 0.035 & \textbf{0.83}\\
\cmidrule{1-7}
 & 30 & \textbf{-0.026} & 0.055 & 0.056 & 0.061 & \textbf{0.91}\\

 & 100 & \textbf{-0.025} & 0.030 & 0.031 & 0.040 & \textbf{0.87}\\

\multirow{-3}{*}{ GEE (exch)} & 150 & \textbf{-0.025} & 0.024 & 0.025 & 0.035 & \textbf{0.83}\\
\bottomrule
\end{tabular}}
\label{simu1}
\end{table}

Table \ref{simu2} displays the simulation results of conditional excursion effects under Scenario (1). ECE, ECE-NonP, and the G-estimator parallel the performances of EMEE, EMEE-NonP, and DR-EMEE-NonP, exhibiting minimal bias and great empirical coverage probabilities due to the correct specification of the treatment effect model by including the moderator $Z_t$. However, GEE methods yield biased results across all settings in Tables \ref{simu1} - \ref{simu2}, with empirical coverage probabilities notably diverging from nominal values.


\begin{table}[ht]
\caption{Comparison of eight estimators of the treatment effect moderation $Z_t$ under Scenario (1)}
\centering
\resizebox{1\textwidth}{!}{%
\begin{tabular}[t]{cccccccccccc}
\toprule
& &   \multicolumn{5}{c}{$\beta_0$}  &   \multicolumn{5}{c}{$\beta_1$}\\
\cmidrule(lr){3-7} \cmidrule(lr){8-12}
Estimator & Time Length & Bias & SE & SD & RMSE & CP & Bias & SE & SD & RMSE & CP\\
\midrule

 & 30 & -0.003 & 0.076 & 0.077 & 0.077 & 0.94 & 0.004 & 0.076 & 0.077 & 0.077 & 0.95\\

 & 100 & 0.001 & 0.041 & 0.043 & 0.043 & 0.94 & -0.002 & 0.041 & 0.043 & 0.043 & 0.92\\

\multirow{-3}{*}{ECE} & 150 & -0.001 & 0.034 & 0.034 & 0.034 & 0.95 & 0.001 & 0.034 & 0.035 & 0.035 & 0.94\\
\cmidrule{1-12}
 & 30 & -0.003 & 0.076 & 0.078 & 0.078 & 0.94 & 0.004 & 0.076 & 0.077 & 0.077 & 0.95\\

 & 100 & 0.001 & 0.042 & 0.043 & 0.043 & 0.95 & -0.002 & 0.042 & 0.043 & 0.043 & 0.93\\

\multirow{-3}{*}{ECE-NonP} & 150 & -0.001 & 0.034 & 0.034 & 0.034 & 0.95 & 0.001 & 0.034 & 0.035 & 0.035 & 0.95\\
\cmidrule{1-12}
 & 30 & -0.003 & 0.076 & 0.077 & 0.077 & 0.94 & 0.004 & 0.076 & 0.077 & 0.077 & 0.95\\

 & 100 & 0.001 & 0.042 & 0.042 & 0.042 & 0.95 & -0.002 & 0.041 & 0.043 & 0.043 & 0.93\\

\multirow{-3}{*}{G-estimator} & 150 & -0.001 & 0.034 & 0.034 & 0.034 & 0.95 & 0.001 & 0.034 & 0.035 & 0.035 & 0.95\\

\cmidrule{1-12}
 & 30 & -0.002 & 0.076 & 0.077 & 0.077 & 0.94 & 0.003 & 0.076 & 0.078 & 0.078 & 0.94\\

 & 100 & 0.001 & 0.041 & 0.042 & 0.042 & 0.94 & -0.002 & 0.041 & 0.043 & 0.043 & 0.93\\

\multirow{-3}{*}{EMEE} & 150 & -0.001 & 0.034 & 0.034 & 0.034 & 0.94 & 0.001 & 0.034 & 0.035 & 0.035 & 0.95\\
\cmidrule{1-12}
 & 30 & -0.003 & 0.076 & 0.077 & 0.077 & 0.94 & 0.003 & 0.076 & 0.078 & 0.078 & 0.95\\

 & 100 & 0.001 & 0.042 & 0.042 & 0.042 & 0.95 & -0.002 & 0.042 & 0.043 & 0.043 & 0.93\\

\multirow{-3}{*}{EMEE-NonP} & 150 & -0.001 & 0.034 & 0.034 & 0.034 & 0.95 & 0.001 & 0.034 & 0.035 & 0.035 & 0.95\\
\cmidrule{1-12}
 & 30 & -0.003 & 0.076 & 0.077 & 0.077 & 0.94 & 0.003 & 0.076 & 0.078 & 0.078 & 0.95\\

 & 100 & 0.001 & 0.042 & 0.042 & 0.042 & 0.95 & -0.002 & 0.042 & 0.043 & 0.043 & 0.93\\

\multirow{-3}{*}{DR-EMEE-NonP} & 150 & -0.001 & 0.034 & 0.034 & 0.034 & 0.95 & 0.001 & 0.034 & 0.035 & 0.035 & 0.95\\
\cmidrule{1-12}
 & 30 & \textbf{0.021} & 0.074 & 0.075 & 0.078 & 0.94 & \textbf{-0.024} & 0.072 & 0.073 & 0.077 & \textbf{0.92}\\

 & 100 & \textbf{0.024} & 0.040 & 0.041 & 0.048 & \textbf{0.89} & \textbf{-0.027} & 0.039 & 0.041 & 0.049 & \textbf{0.88}\\

\multirow{-3}{*}{GEE (ind)} & 150 & \textbf{0.022} & 0.033 & 0.033 & 0.039 & \textbf{0.90} & \textbf{-0.024} & 0.032 & 0.033 & 0.041 & \textbf{0.87}\\
\cmidrule{1-12}
 & 30 & \textbf{0.021} & 0.074 & 0.075 & 0.078 & 0.94 & \textbf{-0.024} & 0.072 & 0.073 & 0.077 & \textbf{0.92}\\

 & 100 & \textbf{0.024} & 0.040 & 0.041 & 0.048 & \textbf{0.89} & \textbf{-0.027} & 0.039 & 0.041 & 0.049 & \textbf{0.88}\\

\multirow{-3}{*}{GEE (exch)} & 150 & \textbf{0.022} & 0.033 & 0.033 & 0.039 & \textbf{0.90} & \textbf{-0.025} & 0.032 & 0.033 & 0.041 & \textbf{0.87}\\
\bottomrule
\end{tabular}}
\label{simu2}
\end{table}

Table \ref{simu3} summarizes simulation results of marginal excursion effects under Scenario (2). Only DR-EMEE-NonP performs well, exhibiting minimal bias and impressive empirical coverage probabilities due to its rate double robustness property lacking in other methods.  Yet, in Table \ref{simu4}, all methods exhibit excellent performances attributed to correct specification of the outcome model and treatment effect model. Scenarios (3) and (4) yield similar observations; details are provided in Appendix \ref{sup7}.

\begin{table}
\caption{Comparison of eight estimators for the fully marginal excursion effect under Scenario (2)}
\centering
\resizebox{0.8\textwidth}{!}{%
\begin{tabular}[t]{ccccccc}
\toprule
Estimator & Time Length & Bias & SE & SD & RMSE & CP\\
\midrule
 & 30 & \textbf{-0.015} & 0.065 & 0.067 & 0.068 & 0.93\\

 & 100 & \textbf{-0.015} & 0.036 & 0.036 & 0.039 & \textbf{0.92}\\

\multirow{-3}{*}{ECE} & 150 & \textbf{-0.015} & 0.029 & 0.030 & 0.034 & \textbf{0.91}\\
\cmidrule{1-7}
 & 30 & \textbf{-0.017} & 0.062 & 0.067 & 0.069 & \textbf{0.92}\\

 & 100 & \textbf{-0.015} & 0.034 & 0.036 & 0.039 & \textbf{0.91}\\

\multirow{-3}{*}{ECE-NonP} & 150 & \textbf{-0.016} & 0.028 & 0.030 & 0.034 & \textbf{0.89}\\
\cmidrule{1-7}
 & 30 & \textbf{-0.015} & 0.065 & 0.067 & 0.068 & 0.93\\

 & 100 & \textbf{-0.015} & 0.036 & 0.036 & 0.039 & \textbf{0.92}\\

\multirow{-3}{*}{G-estimator} & 150 & \textbf{-0.015} & 0.029 & 0.030 & 0.034 & \textbf{0.91}\\
\cmidrule{1-7}
 & 30 & \textbf{-0.015} & 0.065 & 0.067 & 0.068 & 0.93\\

 & 100 & \textbf{-0.015} & 0.036 & 0.036 & 0.039 & \textbf{0.92}\\

\multirow{-3}{*}{EMEE} & 150 & \textbf{-0.015} & 0.029 & 0.030 & 0.034 & \textbf{0.91}\\
\cmidrule{1-7}
 & 30 & \textbf{-0.017} & 0.067 & 0.067 & 0.069 & 0.94\\

 & 100 & \textbf{-0.015} & 0.037 & 0.036 & 0.039 & 0.93\\

\multirow{-3}{*}{EMEE-NonP} & 150 & \textbf{-0.016} & 0.030 & 0.030 & 0.034 & \textbf{0.92}\\
\cmidrule{1-7}
 & 30 & -0.003 & 0.066 & 0.068 & 0.068 & 0.94\\

 & 100 & -0.002 & 0.036 & 0.037 & 0.037 & 0.94\\

\multirow{-3}{*}{DR-EMEE-NonP} & 150 & -0.002 & 0.030 & 0.031 & 0.031 & 0.95\\
\cmidrule{1-7}
 & 30 & \textbf{-0.031} & 0.062 & 0.064 & 0.071 & \textbf{0.92}\\

 & 100 & \textbf{-0.031} & 0.034 & 0.034 & 0.046 & \textbf{0.85}\\

\multirow{-3}{*}{GEE (ind)} & 150 & \textbf{-0.032} & 0.028 & 0.029 & 0.043 & \textbf{0.78}\\
\cmidrule{1-7}
 & 30 & \textbf{-0.031} & 0.062 & 0.064 & 0.071 & \textbf{0.92}\\

 & 100 & \textbf{-0.031} & 0.034 & 0.034 & 0.046 & \textbf{0.85}\\

\multirow{-3}{*}{GEE (exch)} & 150 & \textbf{-0.032} & 0.028 & 0.029 & 0.043 & \textbf{0.78}\\
\bottomrule
\end{tabular}}
\label{simu3}
\end{table}

\begin{table}
\caption{Comparison of eight estimators of the treatment effect moderation $Z_t$ under Scenario (2)}
\centering
\resizebox{1\textwidth}{!}{%
\begin{tabular}[t]{cccccccccccc}
\toprule
& &   \multicolumn{5}{c}{$\beta_0$}  &   \multicolumn{5}{c}{$\beta_1$}\\
\cmidrule(lr){3-7} \cmidrule(lr){8-12}
Estimator & Time Length & Bias & SE & SD & RMSE & CP & Bias & SE & SD & RMSE & CP\\
\midrule
 & 30 & 0.002 & 0.084 & 0.088 & 0.088 & 0.95 & -0.002 & 0.080 & 0.081 & 0.081 & 0.94\\

 & 100 & 0.000 & 0.046 & 0.046 & 0.046 & 0.95 & 0.000 & 0.044 & 0.044 & 0.044 & 0.95\\

\multirow{-3}{*}{ECE} & 150 & -0.001 & 0.038 & 0.038 & 0.038 & 0.95 & 0.001 & 0.036 & 0.035 & 0.035 & 0.94\\
\cmidrule{1-12}
 & 30 & -0.004 & 0.103 & 0.089 & 0.089 & 0.98 & 0.000 & 0.076 & 0.082 & 0.082 & 0.92\\

 & 100 & -0.005 & 0.057 & 0.046 & 0.047 & 0.99 & 0.002 & 0.041 & 0.045 & 0.045 & 0.94\\

\multirow{-3}{*}{ECE-NonP} & 150 & -0.005 & 0.046 & 0.038 & 0.038 & 0.98 & 0.002 & 0.034 & 0.036 & 0.036 & 0.93\\
\cmidrule{1-12}
 & 30 & 0.002 & 0.083 & 0.086 & 0.086 & 0.95 & -0.002 & 0.078 & 0.080 & 0.080 & 0.95\\

 & 100 & 0.000 & 0.045 & 0.045 & 0.045 & 0.95 & 0.000 & 0.043 & 0.043 & 0.043 & 0.95\\

\multirow{-3}{*}{G-estimator} & 150 & -0.001 & 0.037 & 0.037 & 0.037 & 0.95 & 0.001 & 0.035 & 0.034 & 0.034 & 0.94\\
\cmidrule{1-12}
 & 30 & 0.002 & 0.083 & 0.086 & 0.086 & 0.95 & -0.002 & 0.078 & 0.080 & 0.080 & 0.95\\

 & 100 & 0.000 & 0.045 & 0.045 & 0.045 & 0.95 & 0.000 & 0.043 & 0.043 & 0.043 & 0.95\\

\multirow{-3}{*}{EMEE} & 150 & -0.001 & 0.037 & 0.037 & 0.037 & 0.95 & 0.001 & 0.035 & 0.034 & 0.034 & 0.94\\
\cmidrule{1-12}
 & 30 & -0.004 & 0.083 & 0.088 & 0.088 & 0.94 & 0.000 & 0.076 & 0.081 & 0.081 & 0.94\\

 & 100 & -0.005 & 0.045 & 0.046 & 0.046 & 0.95 & 0.002 & 0.042 & 0.044 & 0.044 & 0.94\\

\multirow{-3}{*}{EMEE-NonP} & 150 & -0.005 & 0.037 & 0.038 & 0.038 & 0.94 & 0.002 & 0.034 & 0.035 & 0.035 & 0.94\\
\cmidrule{1-12}
 & 30 & -0.003 & 0.086 & 0.088 & 0.088 & 0.95 & 0.000 & 0.075 & 0.081 & 0.081 & 0.93\\

 & 100 & -0.004 & 0.047 & 0.046 & 0.046 & 0.96 & 0.002 & 0.041 & 0.044 & 0.044 & 0.94\\

\multirow{-3}{*}{DR-EMEE-NonP} & 150 & -0.005 & 0.038 & 0.038 & 0.038 & 0.95 & 0.002 & 0.034 & 0.035 & 0.035 & 0.94\\
\cmidrule{1-12}
 & 30 & 0.002 & 0.085 & 0.089 & 0.089 & 0.94 & -0.002 & 0.080 & 0.082 & 0.082 & 0.94\\

 & 100 & 0.000 & 0.047 & 0.046 & 0.046 & 0.96 & 0.000 & 0.044 & 0.044 & 0.044 & 0.96\\

\multirow{-3}{*}{GEE (ind)} & 150 & -0.001 & 0.038 & 0.038 & 0.038 & 0.95 & 0.001 & 0.036 & 0.035 & 0.035 & 0.95\\
\cmidrule{1-12}
 & 30 & 0.002 & 0.085 & 0.089 & 0.089 & 0.94 & -0.002 & 0.080 & 0.082 & 0.082 & 0.94\\

 & 100 & 0.000 & 0.047 & 0.046 & 0.046 & 0.96 & 0.000 & 0.044 & 0.044 & 0.044 & 0.96\\

\multirow{-3}{*}{GEE (exch)} & 150 & -0.001 & 0.038 & 0.038 & 0.038 & 0.95 & 0.001 & 0.036 & 0.035 & 0.035 & 0.94\\
\bottomrule
\end{tabular}}
\label{simu4}
\end{table}

\section{APPLICATION TO DRINK LESS DATA}
\label{data}
In this section, we analyze the Drink Less trial data to evaluate the efficacy of delivering push notifications on user engagement with the Drink Less app.  Due to the form of intervention, participants were available for the intervention at all times, i.e., $I_t = 1$. We exclusively apply EMEE, EMEE-NonP, and DR-EMEE-NonP to the Drink Less data, focusing solely on marginal causal excursion effects, for which ECE, ECE-NonP, and G-estimator have been identified as potentially biased. For EMEE, we adopt a working model $g(H_t)^{\top}\alpha = \alpha_0 + \alpha_1 Z_t$ for the logarithm of the expected outcome under no treatment. In EMEE-NonP and DR-EMEE-NonP, we leverage two-part generalized additive models to estimate the conditional mean of proximal outcomes. Baseline and time-varying covariates including age, gender, employment type, AUDIT score, days since download, and the number of screen views yesterday are employed as control variables in $g(H_t)$. 

\subsection{Primary Analysis}
In the primary analysis, we estimate the marginal excursion effect of push notifications on the number of screen views. We adopt an analysis model with $S_t = 1$, given by:
\begin{eqnarray*}
    \log \frac{\mathbb{E}\left\{\mathbb{E}\left(Y_{t, 1} \mid H_t, A_t=1\right)\right\}}{\mathbb{E}\left\{\mathbb{E}\left(Y_{t, 1} \mid H_t, A_t=0\right)\right\}}=\beta_0.
\end{eqnarray*}

Results in Table \ref{drink1} show all estimators indicate the effect is statistically significant from zero, with EMEE-NonP and DR-EMEE-NonP yielding marginally reduced standard errors. Subsequently, we compare the marginal excursion effect between the standard notification and the notification from a new message bank using the following analysis model:
\begin{eqnarray*}
    \log \frac{\mathbb{E}\left\{\mathbb{E}\left(Y_{t, 1} \mid H_t, \bmath{A}_t\right)\right\}}{\mathbb{E}\left\{\mathbb{E}\left(Y_{t, 1} \mid H_t,\bmath{A}_t=0\right)\right\}}=\beta_{0,1}A_{t,1} + \beta_{0,2}A_{t,2},
\end{eqnarray*}
where $A_{t,1} = 1$ denotes the standard notification and $A_{t,2}=1$ the new notification. Findings in Table \ref{drink1} affirm the efficacy of both notifications, with the standard notification having a higher treatment effect.

\begin{table}[ht]
\caption{Marginal excursion effects and effect moderation in the Drink Less micro-randomized trial}
\centering
\resizebox{1\textwidth}{!}{%
\begin{tabular}[t]{lccccccccc}
\toprule
&   \multicolumn{4}{c}{$\beta_0$}  &   \multicolumn{4}{c}{$\beta_1$}\\
\cmidrule(lr){2-5} \cmidrule(lr){6-9}
Estimator & Estimate & SE & 95\% CI & p-Value & Estimate & SE & 95\% CI & p-Value\\
\midrule
\multicolumn{9}{l}{\textbf{Marginal excursion effects of providing notifications}} \\
EMEE & 1.110 & 0.123 & (0.869,1.352) & $<$ 0.001 & & & & \\
EMEE-NonP & 1.120 & 0.111 & (0.903,1.337) & $<$ 0.001 & & & & \\
DR-EMEE-NonP & 1.120 & 0.089 & (0.946,1.294) & $<$ 0.001 & & & & \\
\midrule
\multicolumn{9}{l}{\textbf{Marginal excursion effects of providing standard notifications}} \\
EMEE & 1.242 & 0.132 & (0.983,1.500) & $<$ 0.001 & & & & \\
EMEE-NonP & 1.238 & 0.130 & (0.982,1.493) & $<$ 0.001 & & & & \\
DR-EMEE-NonP & 1.240 & 0.100 & (1.042,1.436) & $<$ 0.001 & & & & \\
\midrule
\multicolumn{9}{l}{\textbf{Marginal excursion effects of providing new notifications}} \\
EMEE & 0.965 & 0.138 & (0.694,1.236) & $<$ 0.001 & & & & \\
EMEE-NonP & 0.988 & 0.129 & (0.736,1.240) & $<$ 0.001 & & & & \\
DR-EMEE-NonP & 0.991 & 0.099 & (0.798,1.184) & $<$ 0.001 & & & & \\
\midrule
\multicolumn{9}{l}{\textbf{Effect moderation of days since download}} \\
EMEE & 1.352 & 0.181 & (0.997,1.707) & $<$ 0.001 & -0.019 & 0.011 & (-0.040,0.003) & 0.091\\
EMEE-NonP & 1.555 & 0.185 & (1.187,1.914) & $<$ 0.001 & \textbf{-0.034} & 0.013 & \textbf{(-0.059,-0.009)} & \textbf{0.001}\\
DR-EMEE-NonP & 1.477 & 0.153 & (1.177,1.777) & $<$ 0.001 & \textbf{-0.027} & 0.010 & \textbf{(-0.046,-0.008)} & \textbf{0.001}\\
\midrule
\multicolumn{9}{l}{\textbf{Effect moderation of the number of screen views yesterday}} \\
EMEE & 1.195 & 0.164 & (0.873,1.517) & $<$ 0.001 & -0.023 & 0.024 & (-0.071,0.024) & 0.329\\
EMEE-NonP & 1.116 & 0.122 & (0.878,1.355) & $<$ 0.001 & 0.000 & 0.018 & (-0.034,0.035) & 0.982\\
DR-EMEE-NonP & 1.118 & 0.096 & (0.930,1.307) & $<$ 0.001 & 0.001 & 0.017 & (-0.032,0.033) & 0.974\\
\bottomrule
\end{tabular}}
\label{drink1}
\end{table}

\subsection{Secondary Analysis}
For the secondary analysis, we examine the effect moderation of providing push notifications on user engagement by setting $S_t$ to variables such as ``days since download'' and ``the number of screen views yesterday''. The analysis model is:
\begin{eqnarray*}
    \log \frac{\mathbb{E}\left\{\mathbb{E}\left(Y_{t, 1} \mid H_t, A_t=1\right) \mid S_t\right\}}{\mathbb{E}\left\{\mathbb{E}\left(Y_{t, 1} \mid H_t, A_t=0\right) \mid S_t\right\}}=\beta_0+\beta_1 S_t. 
\end{eqnarray*}

Findings in Table \ref{drink1} show EMEE-NonP and DR-EMEE-NonP identify ``days since download" as a significant moderator, overlooked by EMEE.  Furthermore, all estimators reveal ``number of screen views yesterday" does not significantly influence the treatment effect.

Based on the findings, push notifications significantly affect user engagement, with the standard notification exhibiting a higher effect. Additionally, over prolonged use, users seem to get habituated, resulting in diminishing treatment effects. This highlights the potential utility of periodically refreshing intervention strategies to maintain user engagement.

\section{DISCUSSION}
\label{discuss}
This paper revisits causal excursion effects, contrasting two excursions from the current treatment protocol into the future, focusing on zero-inflated count proximal outcomes.  We extend ECE and EMEE methods from Qian et al. (2021a) - originally developed for binary outcomes - to zero-inflated count outcomes and introduce novel nonparametric estimators for nuisance function estimation. Notably, zero-inflated count outcomes are relatively unexplored in mHealth, with few discussions on extensions to count outcomes lacking theoretical and empirical investigation. We demonstrate the rate double robustness of ECE-NonP and DR-EMEE-NonP estimators, useful for analyzing observational mHealth data. Furthermore, we establish consistency and asymptotic normality for ECE-NonP, EMEE-NonP, and DR-EMEE-NonP under bidirectional asymptotics \citep{yu2023}, requiring either sample size or number of decision points to go to infinity.


We summarize a few directions for future research. Firstly, incorporating random effects into causal excursion effects could enable person-specific effects for informing decision-making. Secondly, identifying potential moderators during initial research phases could guide the selection of $S_t$ from $H_t$, currently governed by researcher judgment.  Lastly, while our approach employs nonparametric methods for estimating nuisance functions, with the causal excursion effect model remaining parametric for interpretability of low-dimensional models, future research could extend into nonparametric methods for the causal excursion effect, thereby introducing enhanced flexibility.


\backmatter


\section*{ACKNOWLEDGEMENTS}
We thank Dr. Claire Garnett, Dr. Olga Perski, Dr. Henry WW Potts, and Dr. Elizabeth Williamson for their important contributions to the Drink Less MRT. 
The authors would also like to thank the anonymous referees, the Associate Editor, and the Editor for their conscientious efforts and constructive comments which improved the quality of this paper. 
\vspace*{-8pt}



\section*{FUNDING}
Xueqing Liu is supported by PhD student scholarship from the Duke-NUS Medical School, Singapore. Lauren Bell is supported by a PhD studentship funded by the MRC Network of Hubs for Trials Methodology Research (MR/L004933/2- R18). Bibhas Chakraborty would like to acknowledge support from the grant MOE-T2EP20122-0013 from the Ministry of Education, Singapore. 
\vspace*{-8pt}

\section*{DATA AVAILABILITY STATEMENT}
The Drink Less MRT data that support the findings of this paper are publicly available at \url{https://osf.io/mtcfa}.


%
\bibliographystyle{biom} 
\bibliography{output.bib}










\label{lastpage}

\clearpage
\begin{appendices}

\section{Extension to Multi-Category Treatment}
\label{multi}
In the main paper, our discussions predominantly focused on binary treatments. Notably, the proposed methods can be readily adapted to scenarios with multiple treatment categories, a case exemplified by the Drink Less study. 

Consider a scenario with $K$ treatment options. To denote each specific treatment option at time $t$, one can employ dummy variables, denoted as $\bmath{A}_t = (A_{t,1},\cdots, A_{t,K})$. Specifically, the configuration where $A_{t,k} = 1$  (and all other elements are $0$) represents the selection of the $k$th treatment. Conversely, a value of $\bmath{A}_t=0$  indicates the absence of any treatment. 

Illustratively, the model for the causal excursion effect can be expressed as:
\begin{eqnarray*}
    \log \frac{\mathbb{E}\left\{\mathbb{E}\left(Y_{t, 1} \mid H_t, \bmath{A}_t\right)\right\}}{\mathbb{E}\left\{\mathbb{E}\left(Y_{t, 1} \mid H_t, \bmath{A}_t=0\right)\right\}}=S_t^{\top}\beta_{1} A_{t,1}+\cdots+S_t^{\top}\beta_{K} A_{t,K}.
\end{eqnarray*}
Within this equation, $(\beta_{1},\cdots, \beta_{K})$ represents the set of parameter vectors yet to be determined, each denoting the causal excursion effect of the corresponding treatment. 

When transitioning to multi-treatment settings, certain modifications to the estimating equation are imperative. The modification of DR-EMEE-NonP and EMEE-NonP involves representing the nuisance parameter $h(H_t)$ with $K$ conditional means, which is
\begin{eqnarray*}
      \mathbb{E}[U_t(\beta)|H_t, I_t=1] &=& \mu_0\left(1-\widetilde{p}_{t,1}(S_t)-\cdots - \widetilde{p}_{t,K}(S_t)\right) \\
      &+& \mu_1 \exp\{-S_t^{\top}\beta_1\}\widetilde{p}_{t,1}(S_t) + \cdots + \mu_K\exp\{-S_t^{\top}\beta_K\}\widetilde{p}_{t,K}(S_t),
\end{eqnarray*}
where $\mu_{k} = \mathbb{E}[Y_{t,1}|A_{t,1}=0,\cdots, A_{t,k}=1,\cdots,A_{t,K}=0,H_t, I_t=1]$ for $k = 1,\cdots,K$. In addition, the weight $W_t$ should be modified as
\begin{eqnarray*}
    \left\{\frac{\widetilde{p}_{t,1}\left(S_t\right)}{p_{t,1}\left(H_t\right)}\right\}^{A_{t,1}} \left\{\frac{\widetilde{p}_{t,2}\left(S_t\right)}{p_{t,2}\left(H_t\right)}\right\}^{A_{t,2}} \cdots \left\{\frac{1-\widetilde{p}_{t,1}\left(S_t\right)-\cdots-\widetilde{p}_{t,K}\left(S_t\right)}{1-p_{t,1}\left(H_t\right)-\cdots-p_{t,K}\left(H_t\right)}\right\}^{1-A_{t,1}-\cdots-A_{t,K}}
\end{eqnarray*}
Here, $\widetilde{p}_{t,k}$ denotes the numerical probability for each treatment $k$, and $p_{t,k}$ denotes the randomization probability specified by the treatment protocol corresponding to treatment $k$. Other estimators can be similarly modified, we thus omit the details. 
It is crucial to note that reliable estimation of these parameters requires sufficient data, either a large sample size or a sufficiently long follow-up time. 

\section{Derivation of Identifiability Result}
\label{sup1}
This section provides the identifiability result of the causal excursion effect under Assumptions 1 - 3 for $\Delta > 1$. 

\begin{proof}
We first assume that the following statement is true:  

For any $1\leq k \leq \Delta$,
\begin{align}  
\label{state1}
\mathbb{E}\left\{Y_{t, \Delta}\left(\bar{A}_{t-1}, a, \bar{0}\right) \mid H_t, A_t=a, I_t=1\right\}  =  \mathbb{E}\left\{\prod_{j=t+1}^{t+k-1} \frac{\mathbbm{1}\left(A_j=0\right)}{1-p_j\left(H_j\right)} Y_{t, \Delta} \mid H_t, A_t=a, I_t=1\right\}.
\end{align}
Notice that
\begin{align*}
& \mathbb{E}\left\{Y_{t, \Delta}\left(\bar{A}_{t-1}, a, \overline{0}\right) \mid S_t\left(\bar{A}_{t-1}\right), I_t\left(\bar{A}_{t-1}\right)=1\right\} \\
= & \mathbb{E}\left[\mathbb{E}\left\{Y_{t, \Delta}\left(\bar{A}_{t-1}, a, \overline{0}\right) \mid H_t\left(\bar{A}_{t-1}\right), I_t\left(\bar{A}_{t-1}\right)=1\right\} \mid S_t\left(\bar{A}_{t-1}\right), I_t\left(\bar{A}_{t-1}\right)=1\right] \\
= & \mathbb{E}\left[\mathbb{E}\left\{Y_{t, \Delta}\left(\bar{A}_{t-1}, a, \overline{0}\right) \mid H_t, I_t=1\right\} \mid S_t, I_t=1\right] \\
= & \mathbb{E}\left[\mathbb{E}\left\{Y_{t, \Delta}\left(\bar{A}_{t-1}, a, \overline{0}\right) \mid H_t, A_t=a, I_t=1\right\} \mid S_t, I_t=1\right]\\
= & \mathbb{E}\left[\mathbb{E}\left\{\prod_{j=t+1}^{t+\Delta-1} \frac{1\left(A_j=0\right)}{1-p_j\left(H_j\right)} Y_{t, \Delta} \mid A_t=a, H_t, I_t=1\right\} \mid S_t, I_t=1\right],
\end{align*}
where the first equality follows from the law of iterated expectation, the second equality follows from consistency (Assumption 1), the third equality follows from sequential ignorability (Assumption 3), and the last one follows from statement \ref{state1}. 

Next, we prove the preceding statement by induction. For $k = 1$, statement \ref{state1} holds as we define $\prod_{j=t+1}^t \frac{\mathbbm{1}\left(A_j=0\right)}{1-p_j\left(H_j\right)}=1$. In what follows we consider the case $\Delta \geq 2$. Suppose that statement \ref{state1} holds for $k = k_0$, $1\leq k_0 \leq \Delta - 1$. Let us denote $\zeta = \prod_{j=t+1}^{t+k_0-1} \frac{\mathbbm{1}\left(A_j=0\right)}{1-p_j\left(H_j\right)} Y_{t, \Delta}$. Notice that
\begin{align*}
& \mathbb{E}\left(\zeta \mid H_{t+k_0}, A_t=a, I_t=1\right) \\
= & \mathbb{E}\left(\zeta \mid H_{t+k_0}, A_t=a, I_t=1\right) \frac{\mathbb{E}\left\{\mathbbm{1}\left(A_{t+k_0}=0\right) \mid H_{t+k_0}, A_t=a, I_t=1\right\}}{1-p_{t+k_0}\left(H_{t+k_0}, A_t=a, I_t=1\right)} \\
= & \mathbb{E}\left\{\zeta \times \frac{\mathbbm{1}\left(A_{t+k_0}=0\right)}{1-p_{t+k_0}\left(H_{t+k_0}, A_t=a, I_t=1\right)} \mid H_{t+k_0}, A_t=a, I_t=1\right\} \\
= & \mathbb{E}\left\{\prod_{j=t+1}^{t+k_0} \frac{\mathbbm{1}\left(A_j=0\right)}{1-p_j\left(H_j\right)} Y_{t, \Delta} \mid H_{t+k_0}, A_t=a, I_t=1\right\},
\end{align*}

Therefore, we have
\begin{align*}
& \mathbb{E}\left\{Y_{t, \Delta}\left(\bar{A}_{t-1}, a, \overline{0}\right) \mid H_t, A_t=a, I_t=1\right\}=\mathbb{E}\left(\zeta \mid H_t, A_t=a, I_t=1\right) \\
= & \mathbb{E}\left\{\prod_{j=t+1}^{t+k_0} \frac{\mathbbm{1}\left(A_j=0\right)}{1-p_j\left(H_j\right)} Y_{t, \Delta}\left(\bar{A}_{t-1}, a, \overline{0}\right) \mid H_t, A_t=a, I_t=1\right\},
\end{align*}
which completes the proof. 

\end{proof}

\section{Proof of Theorem 1}
\label{sup2}
In this section, we derive the bidirectional asymptotic properties of ECE-NonP. We consider a general observational data framework where treatments $A_t$ are not randomly assigned. Building on the three identification assumptions outlined in Section 3, such an observational study can be interpreted as a sequentially randomized experiment. The distinction, however, is in the randomization probabilities $p_t(H_t)$, which remain unknown and necessitate data-driven estimation, as highlighted by \citep{guo2021discussion, robins1986new}. In observational studies, the nuisance functions of ECE-NonP include three components:
\begin{enumerate}
    \item The conditional mean $\mu_1 = \mathbb{E}\left[Y_{t,1} \mid A_t=1, H_t, I_t = 1\right]$ when $A_t=1$,
    \item The conditional mean $\mu_0 = \mathbb{E}\left[Y_{t,1} \mid A_t=0, H_t, I_t = 1\right]$ when $A_t=0$,
    \item The randomization probability $p_t= p_t(H_t)$.
\end{enumerate}
For simplicity in notation, the triplet function is defined as $\eta = (\mu_1, \mu_0, p_t)$, with its true value being $\eta_0 = (\mu_{1,0}, \mu_{0,0}, p_{t,0})$. The estimator of $\eta$ is denoted by $\widehat{\eta} = (\widehat{\mu}_{1}, \widehat{\mu}_{0}, \widehat{p}_t)$. In what follows, we will show that, given the convergence rates specified for the estimators of the nuisance functions under Assumption 4, both the consistency and asymptotic normality of the estimator $\widehat{\phi}$ can be established. 

Define $\mathbb{P}g\{m_C(H_t; \phi, \eta)\} = \frac{1}{T}\sum_{t=1}^T \mathbb{E}[g\{m_C(H_{t};\phi, \eta)\}]$, where $g$ represents an arbitrary function or operator of $m_C$. Similarly, let $\mathbb{P}_n g\{m_C(H_t;\phi, \eta)\}\allowbreak =\frac{1}{n T} \sum_{i=1}^n \sum_{t=1}^T g\left\{m_C(H_{i,t};\phi, \eta)\right\}$. Here, we assume $T \in \mathbb{N}^{+} \cup\{\infty\}$ and define $T^{-1} \sum_{t=1}^T=\lim_{T \rightarrow \infty} T^{-1} \sum_{t=1}^T$ when $T=\infty$. Euclidean norms are denoted by $\|\cdot\|_2$, while $\mathcal{L}_2(P)$-norms are represented as $\|\cdot\|_{2, P}$, defined as $\|\eta\|_{2, P}^2=$ $T^{-1} \sum_{t=1}^T \int\left\|\eta\left(H_t\right)\right\|_2^2 \mathrm{~d} P\left(H_t\right)$. Consider a function set $\mathcal{G}_{\eta_0}=\left\{\eta:\left\|\eta-\eta_0\right\|_{2, P}<\delta\right\}$ for a given $\delta>0$. The Cartesian product of this set is denoted as $\mathcal{U}=\left\{(\phi, \eta): \phi \in \Theta, \eta \in \mathcal{G}_{\eta_0}\right\}$.

To establish Theorem 1, we require the following regularity conditions.
\begin{assumption}
\label{Sass1}
    The solution to $\mathbb{P} m_C\left(H_t; \phi, \eta_0\right)=0$ is unique. Moreover, if $\left\|\mathbb{P} m_C\left(H_t;\phi_n, \eta_0\right)\right\|_2 \rightarrow 0$, then $\| \phi_n - \phi_0 \|_2 \rightarrow 0$ for any sequence $\left\{\phi_n\right\} \in \Theta$.
\end{assumption}

\begin{assumption}
    \label{Sass2}
    For every $\epsilon>0$, there exists a finite $\epsilon$-net $\mathcal{U}\epsilon$ of $\mathcal{U}$. Furthermore, the set $\mathcal{G}_{\eta_0}$ possesses uniformly integrable entropy, i.e., $\int_0^{\infty} \sup _Q\left\{\log N\left(\epsilon\|F\|_{2, Q}, \mathcal{G}_{\eta_0},\|\cdot\|_{2, Q}\right)\right\}^{1 / 2} \mathrm{~d} \epsilon<\infty$, where $F: \Omega \rightarrow$ $\mathbb{R}^3$ denotes a square-integrable envelope for $\mathcal{G}_{\eta_0}$. The covering number, $N\left(\epsilon, \mathcal{G}_{\eta_0},\|\cdot\|\right)$, represents the minimum number of balls of radius $\epsilon$ needed to cover $\mathcal{G}_{\eta_0}$.
\end{assumption}

\begin{assumption}
    \label{Sass3}
    The following conditions hold:
    \begin{enumerate}
        \item For $t = 1,\ldots, T$, $\left|f(H_t)^{\top}\phi\right|$ is bounded almost surely across all $\phi \in \Theta$.
        \item For $t = 1,\ldots, T$, $\left|Y_{t,1}\right|$, $\left\|f(H_t)\right\|_2$, $\sigma_{0,0}^2\left(H_t\right)$, and $\sigma_{1,0}^2\left(H_t\right)$ are bounded almost surely, where $\sigma_{a, 0}^2\left(H_t\right)=\operatorname{var}\left(Y_t \mid A_t=a, H_t, I_t=1\right)$ for $a \in {0,1}$.
        \item For $t = 1,\ldots, T$, given any $\eta \in \mathcal{G}_{\eta_0}$, there exist some constant $b_0$ such that $\left\|\eta\left(H_t\right)\right\|_2<b_0$ almost surely.
        \item  For $t = 1,\ldots, T$, given any $\phi \in \Theta$ and $\eta \in \mathcal{G}{\eta_0}$, there exists some constant $b^*$ such that $\left\|\partial m_C(H_t;\phi, \eta) / \partial \eta^{\mathrm{T}}\right\|_2=$ $\left\|\left(\partial m_C(H_t;\phi, \eta) / \partial p_t, \partial m_C(H_t;\phi, \eta) / \partial \mu_1, \partial m_C(H_t;\phi, \eta) / \partial \mu_0\right)\right\|_2<b^*$ almost surely.
    \end{enumerate}
\end{assumption}

\begin{assumption}
     \label{Sass4}
     For each $\eta \in \mathcal{G}_{\eta_0}$, 
     \begin{equation*}
         \frac{1}{nT} \sum_{i=1}^n \sum_{t=1}^T \mathbb{E}\{\psi(H_{i,t};\eta)\psi^\top(H_{i,t};\eta)|H_{i,t}\} \rightarrow \Sigma(\eta)
     \end{equation*}
     in probability as $nT \rightarrow \infty$, where $\psi(H_{i,t};\eta) = m_C(H_{i,t};\beta_0,\eta) - \mathbb{E}\{m_C(H_{i,t};\beta_0,\eta)|H_{i,t}\}$ and $\Sigma(\eta)$ is a constant positive-definite matrix for each $\eta$. 
\end{assumption}

\begin{assumption}
    \label{Sass5}
    For $j = 1,\ldots,p$, there exists a constant $M$ such that as $nT \rightarrow \infty$,
    \begin{equation*}
        P\left(\sup_{\eta,\eta'\in \mathcal{G}_{\eta_0}}\frac{(nT)^{-1}\sum_{i=1}^n\sum_{t=1}^T\mathbb{E}[\{\psi^j(H_{i,t};\eta)-\psi^j(H_{i,t};\eta^\prime)\}^2|H_{i,t}]}{\|\eta-\eta^\prime\|_{2,P}^2}\geq M\right) \rightarrow 0,
    \end{equation*}
    where $\psi^j(H_{i,t};\eta)$ represents the $j$th component of $\psi(H_{i,t};\eta)$.
\end{assumption}

We further introduce the subsequent two lemmas for our analysis.

\begin{lemma}
\label{lem1}
    Given that Conditions \ref{Sass2} - \ref{Sass3} are satisfied, as \(n T \rightarrow \infty\), the following holds:
    \begin{align*}
    \sup _{\phi \in \Theta, \eta \in \mathcal{G}_{\eta_0}}\left\|\mathbb{P}_n m_C(H_t;\phi, \eta)-\mathbb{P} m_C(H_t;\phi, \eta)\right\|_2 & \stackrel{P}{\rightarrow} 0, \\
    \sup _{\phi \in \Theta, \eta \in \mathcal{G}_{\eta_0}}\left\|\mathbb{P}_n \dot{m}_\phi(H_t;\phi, \eta)-\mathbb{P} \dot{m}_\phi(H_t;\phi, \eta)\right\|_2 & \stackrel{P}{\rightarrow} 0,
    \end{align*}
    where \(\dot{m}_\phi(H_t;\phi, \eta) = \frac{\partial m_C(H_t;\phi, \eta)}{\partial \phi^{\top}}\).
\end{lemma}

\begin{lemma}
\label{lem2}
Given that Conditions \ref{Sass2} - \ref{Sass5} are satisfied, as \(n T \rightarrow \infty\), the following holds:
\[
\sqrt{n T}\left(\mathbb{P}_n-\mathbb{P}\right) m_C\left(H_t;\phi_0, \eta\right) \rightsquigarrow Z \text { in } l^{\infty}\left(\mathcal{G}_{\eta_0}\right),
\]
where the symbol ``\(\rightsquigarrow\)'' signifies the weak convergence of a stochastic process, and \(l^{\infty}\left(\mathcal{G}_{\eta_0}\right)\) denotes the set of all bounded functions \(f: \mathcal{G}_{\eta_0} \rightarrow \mathbb{R}^p\). The limiting process \(Z=\left\{Z(\eta): \eta \in \mathcal{G}_{\eta_0}\right\}\) is a zero-mean multivariate Gaussian process, with its sample paths belonging to the set \(UC\left(\mathcal{G}_{\eta_0},\|\cdot\|_{2, P}\right) = \left\{z \in l^{\infty}\left(\mathcal{G}_{\eta_0}\right): z\right.\) is uniformly continuous with respect to \(\left.\|\cdot\|_{2, P}\right\}\).
\end{lemma}

The proof of Lemma \ref{lem1} and \ref{lem2} can be found in \citet{yu2023} and is directly applicable to our context. Below we begin the proof of Theorem 1.

\begin{proof}

\textbf{Consistency.}
To establish the consistency of the estimator \(\widehat{\phi}\), we aim to demonstrate that 
\[
\left\|\mathbb{P} m_C\left(H_t;\widehat{\phi}, \eta_0\right)\right\|_2 \stackrel{P}{\rightarrow} 0.
\]

By leveraging the triangle inequality and acknowledging that \(\mathbb{P}_n m_C(H_t; \widehat{\phi}, \widehat{\eta}) = 0\), we obtain the following:
\begin{align}
    \left\|\mathbb{P} m_C\left(H_t;\widehat{\phi}, \eta_0\right)\right\|_2 
    &\leq \left\|\mathbb{P} m_C\left(H_t;\widehat{\phi}, \eta_0\right) - \mathbb{P} m_C(H_t;\widehat{\phi}, \widehat{\eta})\right\|_2 + \left\|\mathbb{P} m_C(H_t;\widehat{\phi}, \widehat{\eta})\right\|_2 \nonumber \\
    &\leq \left\|\mathbb{P} m_C\left(H_t;\widehat{\phi}, \eta_0\right) - \mathbb{P} m_C(H_t;\widehat{\phi}, \widehat{\eta})\right\|_2 \nonumber\\
    &\quad + \sup _{\phi \in \Theta, \eta \in \mathcal{G}_{\eta_0}}\left\|\mathbb{P}_n m_C(H_t;\phi, \eta) - \mathbb{P} m_C(H_t;\phi, \eta)\right\|_2.
    \label{eq2.2}
\end{align}

Now, our goal is to show that the two terms in Equation \eqref{eq2.2} are negligible. Specifically, we must establish that $\left\|\mathbb{P} m_C\left(H_t; \widehat{\phi}, \eta_0\right)-\mathbb{P} m_C(H_t;\widehat{\phi}, \widehat{\eta})\right\|_2 = o_P(1)$ and $\sup _{\phi \in \Theta, \eta \in \mathcal{G}_{\eta_0}}\|\mathbb{P}_n m_C(H_t; \phi, \eta)$ $-\mathbb{P} m_C(H_t; \phi, \eta)\|_2 = o_P(1)$.

Applying Taylor's expansion, we have
\begin{align*}
    \left\|m_C(H_t;\widehat{\phi}, \widehat{\eta})-m_C\left(H_t;\widehat{\phi}, \eta_0\right)\right\|_2 
    &\leq \left\|\left.\frac{\partial m_C(H_t;\phi, \eta)}{\partial \eta^\top}\right|_{\eta=\widetilde{\eta}, \phi=\widehat{\phi}}\right\|_2\left\|\widehat{\eta}-\eta_0\right\|_2,
\end{align*}
where the inequality arises from the Cauchy–Schwarz inequality and the fact that \(\left\|\widetilde{\eta}-\eta_0\right\|_2 < \left\|\widehat{\eta}-\eta_0\right\|_2\). 
Substituting this into the first term of Equation \eqref{eq2.2} yields
\begin{align*}
    \left\|\mathbb{P} m_C(H_t; \widehat{\phi}, \widehat{\eta})-\mathbb{P} m_C\left(H_t; \widehat{\phi}, \eta_0\right)\right\|_2 
    &\leq \left[\frac{1}{T} \sum_{t=1}^T \mathbb{E}\left\{\left\|\left.\frac{\partial m_C(H_t; \phi, \eta)}{\partial \eta^\top}\right|_{\eta=\widetilde{\eta}, \phi=\widehat{\phi}}\right\|_2^2\right\}\right]^{1 / 2} \times \\
    &\phantom{=\ } \left[\frac{1}{T} \sum_{t=1}^T \mathbb{E}\left\{\left\|\widehat{\eta}\left(H_{t}\right)-\eta_0\left(H_{t}\right)\right\|_2^2\right\}\right]^{1 / 2}.
\end{align*}

By Condition \ref{Sass3} and Assumption 4, this further simplifies to
\begin{equation}
    \left\|\mathbb{P} m_C(H_t; \widehat{\phi}, \widehat{\eta})-\mathbb{P} m_C\left(H_t; \widehat{\phi}, \eta_0\right)\right\|_2 \leq b^{\ast}\|\widehat{\eta} - \eta_0\|_{2,P} = o_P(1).
    \label{eq2.3}
\end{equation}

Applying Lemma \ref{lem1} to the second term of Equation \eqref{eq2.2}, we derive
\begin{equation}
    \sup _{\phi \in \Theta, \eta \in \mathcal{G}_{\eta_0}}\left\|\mathbb{P}_n m_C(H_t; \phi, \eta)-\mathbb{P} m_C(H_t; \phi, \eta)\right\|_2 \stackrel{P}{\rightarrow} 0 \text { as } n T \rightarrow \infty.
    \label{eq2.4}
\end{equation}

Equations \eqref{eq2.3} and \eqref{eq2.4}, when plugged into \eqref{eq2.2}, indicate that \(\left\|\mathbb{P} m_C\left(H_t; \widehat{\phi}, \eta_0\right)\right\|_2 = o_P(1)\). Conclusively, Assumption \ref{Sass1} implies \(\widehat{\phi} \stackrel{P}{\rightarrow} \phi_0\) as \(n T \rightarrow \infty\).

\textbf{Asymptotic normality.}
Using Taylor's expansion, we derive
\begin{equation}
\label{eq2.5}
    0 = \sqrt{n T} \mathbb{P}_n m_C(H_t; \widehat{\phi}, \widehat{\eta}) = \sqrt{n T} \mathbb{P}_n m_C\left(H_t; \phi_0, \widehat{\eta}\right) + \mathbb{P}_n \dot{m}_\phi(H_t; \widetilde{\phi}, \widehat{\eta}) \sqrt{n T}\left(\widehat{\phi}-\phi_0\right),
\end{equation}
where the inequality \(\left\|\widetilde{\phi}-\phi_0\right\|_2 < \left\|\widehat{\phi}-\phi_0\right\|_2\) holds. From Lemma \ref{lem1}, it follows that 
\begin{equation*}
\label{eq2.6}
    \sup _{\phi \in \Theta, \eta \in \mathcal{G}_{\eta_0}}\left\|\mathbb{P}_n \dot{m}_\phi(H_t; \phi, \eta)-\mathbb{P} \dot{m}_\phi(H_t; \phi, \eta)\right\|_2 \stackrel{P}{\rightarrow} 0 \text { as } n T \rightarrow \infty.
\end{equation*}

Owing to the consistency of the estimator \(\widehat{\phi}\), \(\widetilde{\phi} \stackrel{P}{\rightarrow} \phi_0\) holds. Additionally, under Assumption 4, the convergence \(\widehat{\eta} \stackrel{P}{\rightarrow} \eta_0\) is guaranteed. Therefore, 
\begin{equation*}
\label{eq2.7}
    \mathbb{P}_n \dot{m}_\phi(H_t; \widetilde{\phi}, \widehat{\eta}) \stackrel{P}{\rightarrow} \mathbb{P} \dot{m}_\phi\left(H_t; \phi_0, \eta_0\right) \text { as } n T \rightarrow \infty.
\end{equation*}

Combining with Equation \eqref{eq2.5}, we have 
\begin{align}
    \sqrt{n T}\left(\widehat{\phi}-\phi_0\right) &= -\left\{\mathbb{P} \dot{m}_\phi\left(H_t; \phi_0, \eta_0\right)\right\}^{-1} \sqrt{n T} \mathbb{P}_n m_C\left(H_t; \phi_0, \widehat{\eta}\right) + o_P(1) \nonumber \\
    &= -\left\{\mathbb{P} \dot{m}_\phi\left(H_t; \phi_0, \eta_0\right)\right\}^{-1} \sqrt{n T} \left\{ \left(\mathbb{P}_n-\mathbb{P}\right) m_C\left(H_t; \phi_0, \widehat{\eta}\right) + \mathbb{P} m_C\left(H_t; \phi_0, \widehat{\eta}\right) \right\} + o_P(1) \label{eq2.8}
\end{align}

Next, we bound the term $\mathbb{P} m_C\left(H_t; \phi_0, \widehat{\eta}\right)$. Expanding this expression, we get:
\begin{align*}
    \mathbb{P} m_C\left(H_t;\phi_0, \widehat{\eta}\right) 
    &= \frac{1}{T} \sum_{t=1}^T \mathbb{E}\biggl[ I_t W_t \biggl( Y_{t,1} \exp \left\{-f(H_t)^{\top}\phi_0 A_{t}\right\} \nonumber 
     -\widehat{\mu}_1\left(H_{t}\right) \exp \left\{f(H_t)^{\top}\phi_0\right\} \widehat{p}_t\left(H_t\right) \nonumber \\
    &\quad -\widehat{\mu}_0\left(H_{t}\right)\left\{1-\widehat{p}_t\left(H_t\right)\right\} \biggr) \times \left\{A_{t}-\widehat{p}_t\left(H_t\right)\right\}f(H_t)\biggr] \nonumber \\
    &= \frac{1}{T} \sum_{t=1}^T \mathbb{E}\biggl[ I_t W_t\biggl(\mu_{1,0}\left(H_{t}\right) \exp \left\{-f(H_t)^{\top}\phi_0\right\} p_{t,0}\left(H_{t}\right) \nonumber 
     \\
     &\quad-\mu_{1,0}\left(H_{t}\right) \exp \left\{-f(H_t)^{\top}\phi_0\right\} \widehat{p}_t\left(H_{t}\right) \nonumber \\
    &\quad +\mu_{1,0}\left(H_{t}\right) \exp \left\{-f(H_t)^{\top}\phi_0\right\} \widehat{p}_t\left(H_{t}\right) \nonumber 
     -\widehat{\mu}_1\left(H_{t}\right) \exp \left\{-f(H_t)^{\top}\phi_0\right\} \widehat{p}_t\left(H_{t}\right) \nonumber \\
    &\quad +\mu_{0,0}\left(H_{t}\right)\left\{1-p_{t,0}\left(H_{t}\right)\right\} \nonumber 
     -\mu_{0,0}\left(H_{t}\right)\left\{1-\widehat{p}_t\left(H_{t}\right)\right\} \nonumber \\
    &\quad +\mu_{0,0}\left(H_{t}\right)\left\{1-\widehat{p}_t\left(H_{t}\right)\right\} \nonumber 
     -\widehat{\mu}_0\left(H_{t}\right)\left\{1-\widehat{p}_t\left(H_{t}\right)\right\} \biggr) \times \left\{p_{t,0}\left(H_{t}\right)-\widehat{p}_t\left(H_{t}\right)\right\}f(H_t)\biggr]
\end{align*}

Utilizing Jensen's inequality and the Cauchy-Schwarz inequality, we can express:
\begin{eqnarray*}
     \|\mathbb{P} m_C\left(H_t;\phi_0, \widehat{\eta}\right)\|_2
     &\leq& \frac{1}{T} \sum_{t=1}^T \mathbb{E}\Big(\Big\|I_t W_t \Big[\mu_{1,0}\left(H_{t}\right) \exp \left\{-f(H_t)^{\top}\phi_0\right\} p_{t,0}\left(H_{t}\right)\\
     &&-\mu_{1,0}\left(H_{ t}\right) \exp \left\{-f(H_t)^{\top}\phi_0 \right\} \widehat{p}_t\left(H_{t}\right)\\
     && +\mu_{1,0}\left(H_{t}\right) \exp \left\{-f(H_t)^{\top}\phi\right\} \widehat{p}_t\left(H_{t}\right)-\widehat{\mu}_1\left(H_{t}\right) \exp \left\{-f(H_t)^{\top}\phi_0 \right\} \widehat{p}_t\left(H_{t}\right) \\
     && +\mu_{0,0}\left(H_{t}\right)\left\{1-p_{t,0}\left(H_{t}\right)\right\}-\mu_{0,0}\left(H_{t}\right)\left\{1-\widehat{p}_t\left(H_{t}\right)\right\}\\
     && +\mu_{0,0}\left(H_{t}\right)\left\{1-\widehat{p}_t\left(H_{t}\right)\right\}-\widehat{\mu}_0\left(H_{t}\right)\left\{1-\widehat{p}_t\left(H_{t}\right)\right\}\Big]\\
     &&\times\left\{p_{t,0}\left(H_{t}\right)-\widehat{p}_t\left(H_{t}\right)\right\}f(H_t) \Big\|_2\Big)\\
     &\leq& \frac{1}{T} \sum_{t=1}^T \mathbb{E}\Big[\left\|f(H_t)\right\|_2 | I_tW_t \mu_{1,0}\left(H_{t}\right) \exp \left\{-f(H_t)^{\top}\phi_0 \right\} p_{t,0}\left(H_{t}\right)\\
     &&-\mu_{1,0}\left(H_{t}\right) \exp \left\{-f(H_t)^{\top}\phi_0 \right\} \widehat{p}_t\left(H_{t}\right)\\
     && +\mu_{1,0}\left(H_{t}\right) \exp \left\{-f(H_t)^{\top}\phi_0 \right\} \widehat{p}_t\left(H_{t}\right)-\widehat{\mu}_1\left(H_{t}\right) \exp \left\{-f(H_t)^{\top}\phi_0 \right\} \widehat{p}_t\left(H_{t}\right)\\
     && +\mu_{0,0}\left(H_{t}\right)\left\{1-p_{t,0}\left(H_{t}\right)\right\}-\mu_{0,0}\left(H_{t}\right)\left\{1-\widehat{p}_t\left(H_{t}\right)\right\}\\
     && +\mu_{0,0}\left(H_{t}\right)\left\{1-\widehat{p}_t\left(H_{t}\right)\right\}-\widehat{\mu}_0\left(H_{t}\right)\left\{1-\widehat{p}_t\left(H_{t}\right)\right\}|\times| p_{t,0}\left(H_{t}\right)-\widehat{p}_t\left(H_{t}\right) \mid\Big]\\
     &\preceq&\frac{1}{T} \sum_{t=1}^T \mathbb{E}\Big[\left\{\left|p_{t,0}\left(H_{t}\right)-\widehat{p}_t\left(H_{t}\right)\right|+\left|\mu_{1,0}\left(H_{t}\right)-\widehat{\mu}_1\left(H_{t}\right)\right|+\left|\mu_{0,0}\left(H_{t}\right)-\widehat{\mu}_0\left(H_{t}\right)\right|\right\}\\
     && \times\left|p_{t,0}\left(H_{t}\right)-\widehat{p}_t\left(H_{t}\right)\right|\Big]\\
     &\preceq& \frac{1}{T} \sum_{t=1}^T\Big\{\Big(\mathbb{E}\big[\big\{\left|p_{t,0}\left(H_{t}\right)-\widehat{p}\left(H_{t}\right)\right|+\left|\mu_{1,0}\left(H_{t}\right)-\widehat{\mu}_1\left(H_{ t}\right)\right|\\
     &&+\left|\mu_{0,0}\left(H_{t}\right)-\widehat{\mu}_0\left(H_{t}\right)\right|\big\}^2\big]\Big)^{1 / 2}\\
     &&\left.\times\left(\mathbb{E}\left[\left\{p_{t,0}\left(H_{t}\right)-\widehat{p}_t\left(H_{t}\right)\right\}^2\right]\right)^{1 / 2}\right\}\\
     &=&\left\|\left|\widehat{p}_t-p_{t,0}\right|+\left|\widehat{\mu}_1-\mu_{1,0}\right|+\left|\widehat{\mu}_0-\mu_{0,0}\right|\right\|_{2, P} \times\left\|\widehat{p}_t-p_{t,0}\right\|_{2, P}\\
     &\leq& \left(\left\|\widehat{p}_t-p_{t,0}\right\|_{2, P}+\left\|\widehat{\mu}_1-\mu_{1,0}\right\|_{2, P}+\left\|\widehat{\mu}_0-\mu_{0,0}\right\|_{2, P}\right)\left\|\widehat{p}_t-p_{t,0}\right\|_{2, P}\\
     &=& o_P\{(nT)^{-1/2}\}
\end{eqnarray*}
where the notation $\preceq$ indicates that the expression on the right serves as an upper bound for the one on the left, and the last inequality follows from Assumption 4. From the preceding derivations, we obtain the following convergence result:
\begin{equation}
\label{eq2.9}
    \sqrt{n T} \mathbb{P} m_C\left(H_t; \phi_0, \widehat{\eta}\right) = o_P(1).
\end{equation} 

In addition, based on Lemma \ref{lem2}, the scaled difference between the empirical measure and the actual probability measure converges weakly as:
\begin{equation*}
    \sqrt{n T}\left(\mathbb{P}_n-\mathbb{P}\right) m_C\left(H_t; \phi_0, \eta\right) \rightsquigarrow Z \quad \text { in } l^{\infty}\left(\mathcal{G}_{\eta_0}\right) \text { as } n T \rightarrow \infty.
\end{equation*}

Given Assumption 4, we know that $\widehat{\eta} \stackrel{P}{\rightarrow} \eta_0$ in the semimetric space $\mathcal{G}_{\eta_0}$ relative to the metric $\|\cdot\|_{2, P}$. As a result, we can establish the joint convergence 
\begin{eqnarray*}
    \left(\sqrt{n T}\left(\mathbb{P}_n-\mathbb{P}\right) m_C\left(H_t; \phi_0, \eta\right), \widehat{\eta}\right) \rightsquigarrow\left(Z, \eta_0\right) \text { in the space } l^{\infty}\left(\mathcal{G}_{\eta_0}\right) \times \mathcal{G}_{\eta_0} \text { as } n T \rightarrow \infty \text {. }
\end{eqnarray*}
where we define a mapping $s: l^{\infty}\left(\mathcal{G}_{\eta_0}\right) \times \mathcal{G}_{\eta_0} \mapsto \mathbb{R}^p$ as $s(z, \eta)=z(\eta)-z\left(\eta_0\right)$. Note that $s$ is continuous at every point $(z,\eta)$ where $\eta \mapsto z(\eta)$ is continuous. Given Lemma \ref{lem2}, we observe that almost all sample paths of $Z$ are continuous on $\mathcal{G}_{\eta_0}$. This further implies that  $s$ is continuous almost everywhere for the points $(Z,\eta_0)$. Utilizing the continuous-mapping theorem, we have
\begin{eqnarray*}
    &&\sqrt{n T}\left(\mathbb{P}_n-\mathbb{P}\right)\left\{m_C\left(H_t; \phi_0, \widehat{\eta}\right)-m_C\left(H_t; \phi_0, \eta_0\right)\right\}\\
    &=&s\left(\sqrt{n T}\left(\mathbb{P}_n-\mathbb{P}\right) m_C\left(H_t; \phi_0, \eta\right), \widehat{\eta}\right) \rightsquigarrow s\left(Z, \eta_0\right)=0.
\end{eqnarray*}

Taking into account Equations \eqref{eq2.8} and \eqref{eq2.9}, we further have
\begin{eqnarray*}
    \sqrt{n T}\left(\widehat{\phi}-\phi_0\right)=-\left\{\mathbb{P} \dot{m}_\phi\left(H_t; \phi_0, \eta_0\right)\right\}^{-1} \sqrt{n T}\left(\mathbb{P}_n-\mathbb{P}\right) m_C\left(H_t; \phi_0, \eta_0\right)+o_P(1)
\end{eqnarray*}

Next, we aim to establish that
\begin{eqnarray}
\label{eq2.10}
    \sqrt{n T}\left(\mathbb{P}_n-\mathbb{P}\right) m_C\left(H_t; \phi_0, \eta_0\right) \stackrel{d}{\rightarrow} M V N(0, \Sigma) \text { as } n T \rightarrow \infty.
\end{eqnarray}
For a given integer satisfying $1 \leq g \leq n T$, define $i(g)$ as the quotient of $g+T$ divided by $T$, and $t(g)$ as the integer  satisfying $g=\{i(g)-1\} T+t(g) \text { and } 1 \leq t(g) \leq T$.
With this setup, proving \eqref{eq2.10} is equivalent to proving
\begin{eqnarray}
    \label{eq2.11}
    \sum_{g=1}^{n T}(n T)^{-1 / 2} \Sigma^{-1 / 2} m_C\left(H_{i(g), t(g)}; \phi_0, \eta_0\right) \stackrel{d}{\rightarrow} M V N(0, I) \text { as } n T \rightarrow \infty.
\end{eqnarray}

Let $\mathcal{F}_0=\left\{L_{1,1}\right\}$. Iteratively, we define the sequence $\left\{\mathcal{F}_g\right\}{1 \leq g \leq n T}$ as
\begin{align*}
\mathcal{F}g &=
\begin{cases}
\mathcal{F}_{g-1} \cup\left\{A_{i(g), t(g)}, Y_{i(g), t(g)}, L_{i(g), t(g)+1}\right\}, & \text{if } t(g)<T, \\
\mathcal{F}_{g-1} \cup\left\{A_{i(g), T}, Y_{i(g), T}, L_{i(g)+1,1}\right\}, & \text{otherwise.}
\end{cases}
\end{align*}
Recall that $\mathbb{E}\left\{m_C\left(H_{i(g), t(g)}; \phi_0, \eta_0\right) \mid \mathcal{F}_{g-1}\right\}=0$. Consequently, the left-hand side of \eqref{eq2.11} comprises a martingale difference sequence with respect to the filtration ${\sigma(\mathcal{F}_g)}_{g\geq 1}$, where $\sigma(\mathcal{F}_g)$ denotes the $\sigma$-algebra generated by $\mathcal{F}_g$.

To demonstrate the asymptotic normality, we apply the martingale central limit theorem for triangular arrays, specifically, Theorem 2.3 of \citet{Mcleish1974}. We need only to validate the following two conditions:
\begin{enumerate}
    \item $\max _{1 \leq g \leq n T}\left\|(n T)^{-1 / 2} \Sigma^{-1 / 2} m_C\left(H_{i(g), t(g)}; \phi_0, \eta_0\right)\right\|_2 \stackrel{P}{\rightarrow} 0$ as $n T \rightarrow \infty$.
    \item $\frac{1}{n T} \sum_{g=1}^{n T} \Sigma^{-1 / 2} m_C\left(H_{i(g), t(g)}; \phi_0, \eta_0\right) m_C^{\top}\left(H_{i(g), t(g)}; \phi_0, \eta_0\right)\left(\Sigma^{-1 / 2}\right)^\top \stackrel{P}{\rightarrow} I$ as $n T \rightarrow \infty$.
\end{enumerate}

To verify Condition 1, notice that
\begin{eqnarray*}
    &&\max _{1 \leq g \leq n T}\left\|(n T)^{-1 / 2} \Sigma^{-1 / 2} m_C\left(H_{i(g), t(g)}; \phi_0, \eta_0\right)\right\|_2\\
    &\leq& (n T)^{-1 / 2}\left\|\Sigma^{-1 / 2}\right\|_2 \max _{1 \leq g \leq n T}\left\|m_C\left(H_{i(g), t(g)}; \phi_0, \eta_0\right)\right\|_2\\
    &\leq& (n T)^{-1 / 2}\left\|\Sigma^{-1 / 2}\right\|_2 \max _{1 \leq g \leq n T}\bigg\{\left|A_{i(g), t(g)}-p_{t,0}\left(H_{i(g), t(g)}\right)\right| \\
    &&\times \big|I_t\widetilde{K}_t Y_{i(g), t(g)} \exp \left\{-f(H_{i(g), t(g)})^{\top}\phi_0 A_{i(g), t(g)}\right\} \\
    &&-\mu_{1,0}\left(H_{i(g), t(g)}\right) \exp \left\{-f(H_{i(g), t(g)})^{\top}\phi\right\} p_{t,0}\left(H_{i(g), t(g)}\right) -\mu_{0,0}\left(H_{i(g), t(g)}\right)\left\{1-p_{t,0}\left(H_{i(g), t(g)}\right)\right\} \big|\bigg\}
\end{eqnarray*}
According to Assumption \ref{Sass3}, we have $\max _{1 \leq g \leq n T}\left\|(n T)^{-1 / 2} \Sigma^{-1 / 2}m_C\left(H_{i(g), t(g)}; \phi_0, \eta_0\right)\right\|_2 = O_P((nT )^{-1/2})$. Hence, Condition 1 holds. 

To verify Condition 2, observe that
\begin{eqnarray*}
    && \left\|(n T)^{-1} \sum_{g=1}^{n T} \Sigma^{-1 / 2} m_C\left(H_{i(g), t(g)}; \phi_0, \eta_0\right) m_C^{\top}\left(H_{i(g), t(g)}; \phi_0, \eta_0\right)\left(\Sigma^{-1 / 2}\right)^\top-I\right\|_2\\
    &=& \left\|\Sigma^{-1 / 2}\left\{(n T)^{-1} \sum_{g=1}^{n T} m_C\left(H_{i(g), t(g)}; \phi_0, \eta_0\right) m_C^{\top}\left(H_{i(g), t(g)}; \phi_0, \eta_0\right)-\Sigma\right\}\left(\Sigma^{-1 / 2}\right)^\top\right\|_2\\
    &\leq& \left\|\Sigma^{-1 / 2}\right\|_2^2\left\|(n T)^{-1} \sum_{g=1}^{n T} m_C\left(H_{i(g), t(g)}; \phi_0, \eta_0\right) m_C^{\top}\left(H_{i(g), t(g)}; \phi_0, \eta_0\right)-\Sigma\right\|_2.
\end{eqnarray*}
It suffices to show 
\begin{eqnarray*}
    \left\|(n T)^{-1} \sum_{g=1}^{n T} m_C^{\top}\left(H_{i(g), t(g)}; \phi_0, \eta_0\right) m_C^{\top}\left(H_{i(g), t(g)}; \phi_0, \eta_0\right)-\Sigma\right\|_2 = o_P(1).
\end{eqnarray*}
Let us define 
\begin{align*}
    M_g &= m_C\left(H_{i(g), t(g)}; \phi_0, \eta_0\right) m_C^{\top}\left(H_{i(g), t(g)}; \phi_0, \eta_0\right)\\
    &\quad -\mathbb{E}\left\{m_C\left(H_{i(g), t(g)}; \phi_0, \eta_0\right)m_C^{\top}\left(H_{i(g), t(g)}; \phi_0, \eta_0\right) \mid \mathcal{F}_{g-1}\right\}.
\end{align*}
The sequence $\left\{M_g\right\}_{g \geq 1}$ forms a martingale difference sequence with respect to the filtration $\left\{\sigma\left(\mathcal{F}_g\right)\right\}_{g \geq 1}$. Given that $\mathbb{E}\left(M_g M_{g^{\prime}}^T\right)=0$ for $g \neq g^{\prime}$ and $\mathbb{E}\left(M_g M_g^T\right)$ is bounded for all $g$, we have $\left\|(n T)^{-1} \sum_{g=1}^{n T} M_g\right\|_2 \stackrel{P}{\rightarrow} 0$ as $n T \rightarrow \infty$. This leads us to
\begin{eqnarray*}
    &&\Bigg\|\frac{1}{n T} \sum_{g=1}^{n T}\Big[m_C\left(H_{i(g), t(g)}; \phi_0, \eta_0\right) m_C^{\top}\left(H_{i(g), t(g)}; \phi_0, \eta_0\right)\\
    &&-\mathbb{E}\left\{m_C\left(H_{i(g), t(g)}; \phi_0, \eta_0\right) m_C^{\top}\left(H_{i(g), t(g)}; \phi_0, \eta_0\right) \mid \mathcal{F}_{g-1}\right\}\Big]\Bigg\|_2 \stackrel{P}{\rightarrow} 0,
\end{eqnarray*}
which proves Condition 2. Hence, Theorem 1 is established.
\end{proof}

\section{Proof of Theorem 2}
\label{sup3}
In the context of the marginal excursion effect, the necessary conditions and lemmas parallel those previously mentioned, specifically Conditions \ref{Sass1}-\ref{Sass5} and Lemmas \ref{lem1} and \ref{lem2}.  The primary difference involves replacing the estimating function $m_C(H_t;\phi,\eta)$ with $m_M(H_t;\beta,\eta)$,  and substituting $f(H_t)$ with $S_t$.
The proof of consistency in Theorem 2 is the same as that of Theorem 1. When deducing asymptotic normality, we need only to bound the term $\mathbb{P}m_M(H_t;\beta_0, \widehat{\eta})$. The subsequent proof aligns closely with that of Theorem 1 and is thus not reiterated in detail here.

According to the law of iterated expectation, we have
\begin{eqnarray*}
    \mathbb{P}m_M(H_t;\beta_0, \widehat{\eta}) &=& \frac{1}{T}\sum_{t=1}^T \mathbb{E}\Big\{S_t I_t W_t \Big\{Y_{t,1}\exp(-S_t^{\top}\beta_0A_t) - \widehat{\mu}_1(H_t)\exp(-S_t^{\top}\beta_0)\widetilde{p}_t(S_t)\\
    &&-\widehat{\mu}_0(H_t)(1-\widetilde{p}_t(S_t))\Big\}\times \left\{A_t - \widetilde{p}_t(S_t)\right\}\Big\}\\
    &=& \frac{1}{T}\sum_{t=1}^T\mathbb{E}\Big\{\mathbb{E}\Big[S_t I_t W_t\Big\{Y_{t,1}\exp(-S_t^{\top}\beta_0A_t) - \widehat{\mu}_1(H_t)\exp(-S_t^{\top}\beta_0)\widetilde{p}_t(S_t)\\
    &&-\widehat{\mu}_0(H_t)(1-\widetilde{p}_t(S_t))\Big\} \times \left\{A_t - \widetilde{p}_t(S_t)\right\} \mid H_t\Big]
     \Big\}\\
    &=& \frac{1}{T}\sum_{t=1}^T\mathbb{E}\Big\{S_tI_t \mathbb{E}\Big[\Big(Y_{t,1}\exp(-S_t^{\top}\beta_0A_t) - \widehat{\mu}_1(H_t)\exp(-S_t^{\top}\beta_0)\widetilde{p}_t(S_t)\\
    &&-\widehat{\mu}_0(H_t)(1-\widetilde{p}_t(S_t))\Big) \times \left\{A_t - \widetilde{p}_t(S_t)\right\}
    \mid H_t,I_t=1\Big]
     \Big\}\\
    &=& \frac{1}{T}\sum_{t=1}^T\mathbb{E}\Big\{S_tI_t \mathbb{E}\Big[\widetilde{p}_t(S_t)\Big\{Y_{t,1}\exp(-S_t^{\top}\beta_0) - \widehat{\mu}_1(H_t)\exp(-S_t^{\top}\beta_0)\widetilde{p}_t(S_t)\\
    &&-\widehat{\mu}_0(H_t)(1-\widetilde{p}_t(S_t))\Big\} \times \left(1 - \widetilde{p}_t(S_t)\right)
    \mid H_t,I_t=1,A_t=1\Big]
     \Big\} \\
     &&-\mathbb{E}\Big\{S_tI_t \mathbb{E}\Big[\left(1 - \widetilde{p}_t(S_t)\right)\left\{Y_{t,1} - \widehat{\mu}_1(H_t)\exp(-S_t^{\top}\beta_0)\widetilde{p}_t(S_t)-\widehat{\mu}_0(H_t)(1-\widetilde{p}_t(S_t))\right\}\\
    &&\times \widetilde{p}_t(S_t)
    \mid H_t,I_t=1,A_t=0 \Big]
     \Big\} \\
     &=& \frac{1}{T}\sum_{t=1}^T\mathbb{E}\Big\{S_tI_t\widetilde{p}_t(S_t)\left(1 - \widetilde{p}_t(S_t)\right) \left[\exp(-S_t^{\top}\beta_0)\mu_{1,0}-\mu_{0,0}\right]\Big\}\\
     &=& 0
\end{eqnarray*}
where the last equality follows from Model (4.9). 

\section{Proof of Theorem 3}
\label{sup4}
In this section, we provide the proof for Theorem 3, i.e., the bidirectional asymptotics for DR-EMEE-NonP. Again, we omit the proof for consistency. For deriving asymptotic normality, we  need only to bound the term $\mathbb{P}m_D(H_t; \beta_0, \widehat{\eta})$, and the remaining proof is similar to that of Theorem 1.

\begin{eqnarray*}
    \mathbb{P}m_D(H_t;\beta_0, \widehat{\eta}) &=& \frac{1}{T}\sum_{t=1}^T\mathbb{E}\Big\{S_tI_tW_t \left\{Y_{t,1}\exp(-A_tS_t^{\top}\beta_0) - \widehat{\mu}_{A_t}(H_t)\exp(-A_tS_t^{\top}\beta_0)\right\}\times \left\{A_t - \widetilde{p}_t(S_t)\right\} \\
    &&+ S_tI_t\widetilde{p}_t(S_t)(1-\widetilde{p}_t(S_t))\{\exp(-S_t^\top\beta)\widehat{\mu}_1(H_t) - \widehat{\mu}_0(H_t)\}\Big\}\\
    &=& \frac{1}{T}\sum_{t=1}^T\mathbb{E}\Big\{S_tI_tW_t\left\{Y_{t,1}\exp(-A_tS_t^\top\beta_0)-\mu_{A_t,0}(H_t)\exp(-A_tS_t^\top\beta_0)\right\}\times\{A_t-\widetilde{p}_t(S_t)\}\\
    &&+ S_tI_tW_t\left\{\mu_{A_t,0}(H_t)\exp(-A_tS_t^\top\beta_0) - \widehat{\mu}_{A_t}(H_t)\exp(-A_tS_t^\top\beta_0)\right\}\times\{A_t-\widetilde{p}_t(S_t)\}\\
    &&+ S_tI_t\widetilde{p}_t(S_t)(1-\widetilde{p}_t(S_t))\{\widehat{\mu}_{1}(H_t)\exp(-S_t^\top\beta_0)-\widehat{\mu}_{0}(H_t)\}\Big\}\\
    &=& \frac{1}{T}\sum_{t=1}^T\mathbb{E}\Big\{S_tI_t\widetilde{p}_t(S_t)(1-\widetilde{p}_t(S_t))p_{t,0}(H_t)\left(\frac{1}{\widehat{p}_t(H_t)} - \frac{1}{p_{t,0}(H_t)}  + \frac{1}{p_{t,0}(H_t)} \right)\times\\
    &&\left(\mu_{1,0}(H_t)\exp(-S_t^\top\beta_0) - \widehat{\mu}_{1}(H_t)\exp(-S_t^\top\beta_0)\right)\\
    &&- S_tI_t\widetilde{p}_t(S_t)(1-\widetilde{p}_t(S_t))(1-p_{t,0}(H_t))\left(\frac{1}{1-\widehat{p}_t(H_t)} - \frac{1}{1-p_{t,0}(H_t)}  + \frac{1}{1-p_{t,0}(H_t)} \right)\\
    &&\left(\mu_{0,0}(H_t)- \widehat{\mu}_{0}(H_t)\right) + I_tS_t\widetilde{p}_t(S_t)(1-\widetilde{p}_t(S_t))\{\widehat{\mu}_{1}(H_t)\exp(-S_t^\top\beta_0)-\widehat{\mu}_{0}(H_t)\}\Big\}\\
    &=& \frac{1}{T}\sum_{t=1}^T\mathbb{E}\Big\{S_tI_t\widetilde{p}_t(S_t)(1-\widetilde{p}_t(S_t))p_{t,0}(H_t)\left(\frac{1}{\widehat{p}_t(H_t)} - \frac{1}{p_{t,0}(H_t)}  \right)\\
    &&\left(\mu_{1,0}(H_t)\exp(-S_t^\top\beta_0) - \widehat{\mu}_{1}(H_t)\exp(-S_t^\top\beta_0)\right)\\
    &&- S_tI_t\widetilde{p}_t(S_t)(1-\widetilde{p}_t(S_t))(1-p_{t,0}(H_t))\left(\frac{1}{1-\widehat{p}_t(H_t)} - \frac{1}{1-p_{t,0}(H_t)} \right)\\
    && \left(\mu_{0,0}(H_t)- \widehat{\mu}_{0}(H_t)\right)\\
    &&+ S_tI_t\widetilde{p}_t(S_t)(1-\widetilde{p}_t(S_t))\left(\mu_{1,0}(H_t)\exp(-S_t^\top\beta_0) - \widehat{\mu}_{1}(H_t)\exp(-S_t^\top\beta_0)\right)\\
    &&- S_tI_t\widetilde{p}_t(S_t)(1-\widetilde{p}_t(S_t))\left(\mu_{0,0}(H_t)- \widehat{\mu}_{0}(H_t)\right)\\
    &&+S_tI_t\widetilde{p}_t(S_t)(1-\widetilde{p}_t(S_t))\{\widehat{\mu}_{1}(H_t)\exp(-S_t^\top\beta_0)-\widehat{\mu}_{0}(H_t)\}\Big\}\\
    &=& \frac{1}{T}\sum_{t=1}^T\mathbb{E}\Big\{S_tI_t\widetilde{p}_t(S_t)(1-\widetilde{p}_t(S_t))p_{t,0}(H_t)\left(\frac{1}{\widehat{p}_t(H_t)} - \frac{1}{p_{t,0}(H_t)}  \right)\\
    &&\left(\mu_{1,0}(H_t)\exp(-S_t^\top\beta_0) - \widehat{\mu}_{1}(H_t)\exp(-S_t^\top\beta_0)\right)\\
    &&- S_tI_t\widetilde{p}_t(S_t)(1-\widetilde{p}_t(S_t))(1-p_{t,0}(H_t))\left(\frac{1}{1-\widehat{p}_t(H_t)} - \frac{1}{1-p_{t,0}(H_t)} \right)\\
    &&\left(\mu_{0,0}(H_t)- \widehat{\mu}_{0}(H_t)\right)\\
    &&+  S_tI_t\widetilde{p}_t(S_t)(1-\widetilde{p}_t(S_t))\left\{\mu_{1,0}(H_t)\exp(-S_t^\top\beta_0) - \mu_{0,0}(H_t)\right\} \Big\}\\
    &=& \frac{1}{T}\sum_{t=1}^T\mathbb{E}\Big\{S_tI_t\widetilde{p}_t(S_t)(1-\widetilde{p}_t(S_t))p_{t,0}(H_t)\left(\frac{1}{\widehat{p}_t(H_t)} - \frac{1}{p_{t,0}(H_t)}  \right)\\
    &&\left(\mu_{1,0}(H_t)\exp(-S_t^\top\beta_0) - \widehat{\mu}_{1}(H_t)\exp(-S_t^\top\beta_0)\right)\\
    &&-S_tI_t\widetilde{p}_t(S_t)(1-\widetilde{p}_t(S_t))(1-p_{t,0}(H_t))\left(\frac{1}{1-\widehat{p}_t(H_t)} - \frac{1}{1-p_{t,0}(H_t)} \right)\\
    &&\left(\mu_{0,0}(H_t)- \widehat{\mu}_{0}(H_t)\right) \Big\}\\
    &=& \frac{1}{T}\sum_{t=1}^T\mathbb{E}\Big\{S_tI_t\widetilde{p}_t(S_t)\frac{1-\widetilde{p}_t(S_t)}{\widehat{p}_t(H_t)(1-\widehat{p}_t(H_t))}(p_{t,0}(H_t) - \widehat{p}_t(H_t)) \\
    &&\times\big\{(1-\widehat{p}_t(H_t))\mu_{1,0}(H_t)\exp(-S_t^\top\beta_0) - (1-\widehat{p}_t(H_t))\widehat{\mu}_1(H_t)\exp(-S_t^\top\beta_0)\\
    &&+ \widehat{p}_t(H_t)\mu_{0,0}(H_t) - \widehat{p}_t(H_t)\widehat{\mu}_0(H_t)\big\} \Big\}
\end{eqnarray*}
where the third equality follows from $E[Y_{t,1}|H_t,A_t,I_t=1]=\mu_{A_t,0}(H_t)$, the sixth equality follows from Model (4.9). 

Applying Jensen's inequality and Cauchy-Schwarz inequality, we obtain
\begin{eqnarray*}
    \| \mathbb{P}m_D(\beta_0, \widehat{\eta}) \|_2 &\leq& \frac{1}{T}\sum_{t=1}^T\mathbb{E}\Big\{\Big\|I_tS_t\frac{\widetilde{p}_t(S_t)(1-\widetilde{p}_t(S_t))}{\widehat{p}_t(H_t)(1-\widehat{p}_t(H_t))}(p_{t,0}(H_t) - \widehat{p}_t(H_t)) \\
    &\times&\big\{(1-\widehat{p}_t(H_t))\mu_{1,0}(H_t)\exp(-S_t^\top\beta_0) - (1-\widehat{p}_t(H_t))\widehat{\mu}_1(H_t)\exp(-S_t^\top\beta_0)\\
    &&+ \widehat{p}_t(H_t)\mu_{0,0}(H_t) - \widehat{p}_t(H_t)\widehat{\mu}_0(H_t)\big\} \Big\|_2\Big\}\\
    &\leq& \frac{1}{T}\sum_{t=1}^T\mathbb{E} \Big\{\|S_t\|_2 \left|I_tS_t\frac{\widetilde{p}_t(S_t)(1-\widetilde{p}_t(S_t))}{\widehat{p}_t(H_t)(1-\widehat{p}_t(H_t))}(p_{t,0}(H_t) - \widehat{p}_t(H_t))\right| \\
    &\times& \big|\big\{(1-\widehat{p}_t(H_t))\mu_{1,0}(H_t)\exp(-S_t^\top\beta_0) - (1-\widehat{p}_t(H_t))\widehat{\mu}_1(H_t)\exp(-S_t^\top\beta_0)\\
    && + \widehat{p}_t(H_t)\mu_{0,0}(H_t) - \widehat{p}_t(H_t)\widehat{\mu}_0(H_t)\big\} \big|\Big\}\\
    &\preceq& \frac{1}{T}\sum_{t=1}^T\mathbb{E} \left\{|p_{t,0}(H_t) - \widehat{p}_t(H_t)| \times [|\mu_{1,0}(H_t)-\widehat{\mu}_1(H_t)| + |\mu_{0,0}(H_t)-\widehat{\mu}_0(H_t)|]
    \right\}\\
    &\preceq& \frac{1}{T}\sum_{t=1}^T\Big\{\left[\mathbb{E}(|\mu_{1,0}(H_t)-\widehat{\mu}_1(H_t)| + |\mu_{0,0}(H_t)-\widehat{\mu}_0(H_t)|)^2 \right]^{1/2} \\
    &&\times \left[\mathbb{E}(p_{t,0}(H_t) - \widehat{p}_t(H_t))^2 \right]^{1/2}
    \Big\}\\
    &=& \||\mu_{1,0}-\widehat{\mu}_1| + |\mu_{0,0}-\widehat{\mu}_0|\|_{2,P} \times \|p_{t,0} - \widehat{p}_t\|_{2,P}\\
    &\leq& \left(\|\mu_{1,0}-\widehat{\mu}_1\|_{2,P} + \|\mu_{0,0}-\widehat{\mu}_0\|_{2,P}\right)\|p_{t,0} - \widehat{p}_t\|_{2,P} = o_P\{(nT)^{-1/2}\}
\end{eqnarray*}
where the last equality follows from Assumption 5.

\section{Simulation Details}
\label{sup6}
This section presents the simulation settings of Scenarios (3) and (4). 
The data generation procedure of Scenario (3) is the same as that of Scenario (1), except that we adopt the following Thompson sampling algorithm \ref{alg2} to generate the randomization probability \citep{liu2023thompson}.

In Scenario (4), we have two types of treatments and an absence of treatment. Here, we set $\pi_{t} = \exp\left(-0.4(Z_t+0.1)+0.1Z_tA_{t,1}+0.1Z_tA_{t,2}\right)$ and $\mu_{t}=\left\{2.2 \mathbbm{1}_{Z_t=0}+2.5 \mathbbm{1}_{Z_t=1}+2.4 \mathbbm{1}_{Z_t=2}\right\}  \allowbreak e^{A_{t,1}\left(0.1+0.3 Z_t\right) + A_{t,2}\left(0.1+0.1Z_t\right)}$. As a result, the true conditional causal excursion effect is
\begin{eqnarray*}
    \log \frac{\mathbb{E}\left(Y_{t,1} \mid A_t, H_t\right)}{\mathbb{E}\left(Y_{t,1} \mid A_t=0, H_t\right)}=A_{t,1}(0.1+0.4 Z_t) + A_{t,2}(0.1+0.2 Z_t).
\end{eqnarray*}
Further, the fully marginal excursion effect becomes
\begin{eqnarray*}
        \log \frac{\mathbb{E}\left\{\mathbb{E}\left(Y_{t,1} \mid A_t, H_t\right)\right\}}{\mathbb{E}\left\{\mathbb{E}\left(Y_{t,1} \mid A_t=0, H_t\right)\right\}}=0.460A_{t,1} + 0.267A_{t,2}.
\end{eqnarray*}

\begin{algorithm}[t]
\caption{Thompson sampling for count data with normal approximation}
\label{alg2}
\begin{algorithmic}[1]
\REQUIRE Tuning parameter $\alpha = 1$, Initial random exploration time $T_0 = 20$
\STATE \textbf{Initialization:} Randomly choose $A_t \in \mathcal{A}$ for $t\leq T_0$
\FOR{$t=T_0+1,\cdots, T$}
    \STATE Compute the MLE $\widehat{\beta}_{t}$ by maximizing the likelihood function of a Poisson regression
    \STATE Get $\mathcal{I}_t$ as a byproduct of step 3
    \STATE Sample $\widetilde{\beta}_t$ from $\mathcal{N}(\widehat{\beta}_t, \alpha^2\mathcal{I}_t^{-1})$
    \STATE Observe covariates $X_t$ and calculate the probability of $\widehat{p}_t(H_t) = \operatorname{Pr}(f(A_t=1,X_t)^{\top} \widetilde{\beta}_t > f(A_t=0,X_t)^{\top} \widetilde{\beta}_t)$
    \STATE Offer intervention $A_t$ with probability $p_t(H_t) = \max(0.05, \min(0.95, \widehat{p}_t(H_t)))$
\ENDFOR
\end{algorithmic}
\end{algorithm}

\section{Additional Simulation Results}
\label{sup7}
This section presents additional simulation results under Scenario (3) and Scenario (4). For Scenario (3), we employed a Thompson sampling algorithm to determine  $p_t\left( H_t\right)$. Such algorithms have gained popularity in MRTs, facilitating timely and tailored interventions as evidenced in the DIAMANTE study \citep{aguilera2020}. The results closely mirror those of the first scenario, as shown in \ref{simu5} - \ref{simu6}.  Both the EMEE-NonP and EMEE methods yield comparable results, outpacing the GEE methods in performance. Transitioning to Scenario (4), we introduce two types of treatments in addition to an absence of treatment. Similarly, the proposed methods outperform the standard GEE methods, as shown in \ref{simu7} - \ref{simu8}. 


\begin{table}[ht]
\caption{Comparison of five estimators of the fully marginal excursion effect under Scenario (3)}
\centering
\begin{tabular}[t]{ccccccc}
\toprule
Estimator & Time Length & Bias & SE & SD & RMSE & CP\\
\midrule
 & 30 & 0.002 & 0.081 & 0.080 & 0.080 & 0.95\\

 & 100 & -0.003 & 0.061 & 0.061 & 0.061 & 0.96\\

\multirow{-3}{*}{ EMEE} & 150 & 0.005 & 0.053 & 0.054 & 0.054 & 0.94\\
\cmidrule{1-7}
 & 30 & 0.002 & 0.081 & 0.080 & 0.080 & 0.95\\

 & 100 & -0.003 & 0.062 & 0.061 & 0.061 & 0.96\\

\multirow{-3}{*}{ EMEE-NonP} & 150 & 0.005 & 0.053 & 0.054 & 0.054 & 0.95\\
\cmidrule{1-7}
 & 30 & 0.002 & 0.081 & 0.081 & 0.081 & 0.94\\

 & 100 & -0.003 & 0.061 & 0.060 & 0.060 & 0.97\\

\multirow{-3}{*}{ DR-EMEE-NonP} & 150 & 0.005 & 0.053 & 0.053 & 0.053 & 0.94\\
\cmidrule{1-7}
 & 30 & -0.004 & 0.056 & 0.055 & 0.055 & 0.94\\

 & 100 & \textbf{-0.034} & 0.032 & 0.032 & 0.047 & \textbf{0.79}\\

\multirow{-3}{*}{GEE (ind)} & 150 & \textbf{-0.045} & 0.027 & 0.028 & 0.053 & \textbf{0.61}\\
\cmidrule{1-7}
 & 30 & -0.004 & 0.056 & 0.055 & 0.056 & 0.94\\

 & 100 & \textbf{-0.035} & 0.031 & 0.032 & 0.047 & \textbf{0.78}\\

\multirow{-3}{*}{GEE (exch)} & 150 & \textbf{-0.046} & 0.027 & 0.028 & 0.053 & \textbf{0.59}\\
\bottomrule
\end{tabular}
\label{simu5}
\end{table}


\begin{table}[ht]
\caption{Comparison of five estimators of treatment effect moderation $Z_t$ under Scenario (3)}
\centering
\resizebox{1\textwidth}{!}{%
\begin{tabular}[t]{cccccccccccc}
\toprule
& &   \multicolumn{5}{c}{$\beta_0$}  &   \multicolumn{5}{c}{$\beta_1$}\\
\cmidrule(lr){3-7} \cmidrule(lr){8-12}
Estimator & Time Length & Bias & SE & SD & RMSE & CP & Bias & SE & SD & RMSE & CP\\
\midrule
 & 30 & 0.005 & 0.104 & 0.102 & 0.102 & 0.95 & -0.002 & 0.099 & 0.104 & 0.104 & 0.94\\

 & 100 & -0.007 & 0.079 & 0.079 & 0.079 & 0.95 & 0.007 & 0.078 & 0.077 & 0.077 & 0.96\\

\multirow{-3}{*}{ EMEE} & 150 & 0.005 & 0.068 & 0.070 & 0.070 & 0.94 & 0.001 & 0.068 & 0.072 & 0.072 & 0.93\\
\cmidrule{1-12}
 & 30 & 0.006 & 0.104 & 0.102 & 0.103 & 0.95 & -0.002 & 0.100 & 0.104 & 0.104 & 0.95\\

 & 100 & -0.007 & 0.080 & 0.079 & 0.079 & 0.95 & 0.007 & 0.079 & 0.077 & 0.078 & 0.96\\

\multirow{-3}{*}{EMEE-NonP} & 150 & 0.005 & 0.068 & 0.070 & 0.070 & 0.94 & 0.001 & 0.068 & 0.072 & 0.072 & 0.94\\
\cmidrule{1-12}
 & 30 & 0.006 & 0.104 & 0.103 & 0.103 & 0.95 & -0.002 & 0.101 & 0.105 & 0.105 & 0.95\\

 & 100 & -0.007 & 0.080 & 0.079 & 0.079 & 0.96 & 0.008 & 0.079 & 0.077 & 0.078 & 0.96\\

\multirow{-3}{*}{ DR-EMEE-NonP} & 150 & 0.005 & 0.068 & 0.070 & 0.070 & 0.94 & 0.001 & 0.068 & 0.072 & 0.072 & 0.94\\
\cmidrule{1-12}
 & 30 & \textbf{0.018} & 0.073 & 0.073 & 0.075 & 0.95 & \textbf{-0.015} & 0.070 & 0.071 & 0.072 & 0.94\\

 & 100 & \textbf{0.020} & 0.040 & 0.041 & 0.045 & \textbf{0.91} & \textbf{-0.017} & 0.041 & 0.042 & 0.045 & \textbf{0.92}\\

\multirow{-3}{*}{ GEE (ind)} & 150 & \textbf{0.026} & 0.033 & 0.034 & 0.043 & \textbf{0.86} & \textbf{-0.020} & 0.035 & 0.036 & 0.041 & \textbf{0.89}\\
\cmidrule{1-12}
 & 30 & \textbf{0.018} & 0.073 & 0.073 & 0.075 & 0.95 & \textbf{-0.015} & 0.070 & 0.071 & 0.072 & 0.94\\

 & 100 &\textbf{ 0.020} & 0.040 & 0.041 & 0.045 & \textbf{0.92} & \textbf{-0.017} & 0.041 & 0.042 & 0.045 & \textbf{0.92}\\

\multirow{-3}{*}{GEE (exch)} & 150 & \textbf{0.026} & 0.033 & 0.034 & 0.043 & \textbf{0.86} & \textbf{-0.020} & 0.035 & 0.036 & 0.041 & \textbf{0.89}\\
\bottomrule
\end{tabular}}
\label{simu6}
\end{table}


\begin{table}[ht]
\caption{Comparison of five estimators of the fully marginal excursion effects under Scenario (4)}
\centering
\resizebox{1\textwidth}{!}{%
\begin{tabular}[t]{cccccccccccc}
\toprule
& &   \multicolumn{5}{c}{Treatment 1}  &   \multicolumn{5}{c}{Treatment 2}\\
\cmidrule(lr){3-7} \cmidrule(lr){8-12}
Estimator & Time Length & Bias & SE & SD & RMSE & CP & Bias & SE & SD & RMSE & CP\\
\midrule
 & 30 & -0.002 & 0.069 & 0.068 & 0.068 & 0.95 & 0.001 & 0.068 & 0.067 & 0.067 & 0.96\\

 & 100 & -0.001 & 0.037 & 0.038 & 0.038 & 0.94 & 0.001 & 0.037 & 0.039 & 0.039 & 0.94\\

\multirow{-3}{*}{EMEE} & 150 & -0.002 & 0.031 & 0.031 & 0.031 & 0.94 & -0.001 & 0.031 & 0.032 & 0.032 & 0.94\\
\cmidrule{1-12}
 & 30 & -0.002 & 0.069 & 0.068 & 0.068 & 0.95 & 0.001 & 0.069 & 0.067 & 0.067 & 0.96\\

 & 100 & -0.001 & 0.038 & 0.038 & 0.038 & 0.94 & 0.001 & 0.038 & 0.039 & 0.039 & 0.94\\

\multirow{-3}{*}{EMEE-NonP} & 150 & -0.002 & 0.031 & 0.031 & 0.031 & 0.94 & -0.001 & 0.031 & 0.032 & 0.032 & 0.95\\
\cmidrule{1-12}
 & 30 & -0.002 & 0.060 & 0.068 & 0.068 & 0.92 & 0.001 & 0.060 & 0.067 & 0.066 & 0.92\\

 & 100 & -0.001 & 0.033 & 0.038 & 0.038 & 0.91 & 0.001 & 0.033 & 0.039 & 0.039 & 0.91\\

\multirow{-3}{*}{DR-EMEE-NonP} & 150 & -0.002 & 0.027 & 0.031 & 0.031 & 0.91 & -0.001 & 0.027 & 0.032 & 0.032 & 0.92\\
\cmidrule{1-12}
 & 30 & \textbf{-0.020} & 0.066 & 0.066 & 0.069 & 0.94 & \textbf{-0.041} & 0.067 & 0.065 & 0.077 & \textbf{0.92}\\

 & 100 & \textbf{-0.018} & 0.036 & 0.037 & 0.041 & \textbf{0.91} & \textbf{-0.041} & 0.037 & 0.038 & 0.056 & \textbf{0.78}\\

\multirow{-3}{*}{GEE (ind)} & 150 & \textbf{-0.019} & 0.029 & 0.030 & 0.036 & \textbf{0.88} & \textbf{-0.044} & 0.030 & 0.031 & 0.053 & \textbf{0.68}\\
\cmidrule{1-12}
 & 30 & \textbf{-0.020} & 0.066 & 0.066 & 0.069 & 0.94 & \textbf{-0.041} & 0.067 & 0.065 & 0.077 & \textbf{0.91}\\

 & 100 & \textbf{-0.018} & 0.036 & 0.037 & 0.041 & \textbf{0.91} & \textbf{-0.041} & 0.037 & 0.038 & 0.056 & \textbf{0.78}\\

\multirow{-3}{*}{ GEE (exch)} & 150 & \textbf{-0.019} & 0.029 & 0.030 & 0.036 & \textbf{0.88} & \textbf{-0.044} & 0.030 & 0.031 & 0.053 & \textbf{0.69}\\
\bottomrule
\end{tabular}}
\label{simu7}
\end{table}


\begin{table}[ht]
\caption{Comparison of five estimators of treatment effect moderation $Z_t$ under Scenario (4)}
\centering
\resizebox{1\textwidth}{!}{%
\begin{tabular}[t]{cccccccccccc}
\toprule
& &   \multicolumn{5}{c}{$\beta_0$}  &   \multicolumn{5}{c}{$\beta_1$}\\
\cmidrule(lr){3-7} \cmidrule(lr){8-12}
Estimator & Time Length & Bias & SE & SD & RMSE & CP & Bias & SE& SD & RMSE & CP\\
\midrule
\multicolumn{12}{c}{Treatment 1} \\
\cmidrule(lr){1-12}
 & 30 & 0.000 & 0.091 & 0.090 & 0.090 & 0.94 & -0.002 & 0.091 & 0.092 & 0.092 & 0.95\\

 & 100 & 0.000 & 0.050 & 0.049 & 0.049 & 0.95 & 0.000 & 0.050 & 0.049 & 0.049 & 0.96\\

\multirow{-3}{*}{EMEE} & 150 & -0.003 & 0.041 & 0.042 & 0.042 & 0.94 & 0.002 & 0.041 & 0.041 & 0.041 & 0.95\\
\cmidrule{1-12}
 & 30 & 0.000 & 0.092 & 0.090 & 0.090 & 0.95 & -0.002 & 0.091 & 0.092 & 0.092 & 0.95\\

 & 100 & 0.000 & 0.050 & 0.049 & 0.049 & 0.96 & 0.000 & 0.050 & 0.049 & 0.049 & 0.96\\

\multirow{-3}{*}{EMEE-NonP} & 150 & -0.003 & 0.041 & 0.042 & 0.042 & 0.94 & 0.002 & 0.040 & 0.041 & 0.041 & 0.95\\
\cmidrule{1-12}
 & 30 & 0.004 & 0.085 & 0.090 & 0.090 & 0.93 & -0.007 & 0.076 & 0.091 & 0.091 & 0.90\\

 & 100 & 0.003 & 0.046 & 0.049 & 0.049 & 0.94 & -0.005 & 0.042 & 0.048 & 0.049 & 0.91\\

\multirow{-3}{*}{DR-EMEE-NonP} & 150 & 0.000 & 0.038 & 0.042 & 0.042 & 0.93 & -0.002 & 0.034 & 0.041 & 0.041 & 0.89\\
\cmidrule{1-12}
 & 30 & \textbf{0.022} & 0.090 & 0.090 & 0.092 & 0.94 & \textbf{-0.028} & 0.089 & 0.089 & 0.093 & 0.93\\

 & 100 & \textbf{0.023} & 0.049 & 0.048 & 0.053 & 0.93 & \textbf{-0.027} & 0.049 & 0.047 & 0.054 & 0.94\\

\multirow{-3}{*}{GEE (ind)} & 150 & \textbf{0.020} & 0.040 & 0.041 & 0.046 & \textbf{0.92} & \textbf{-0.025} & 0.040 & 0.040 & 0.047 & \textbf{0.90}\\
\cmidrule{1-12}
 & 30 & \textbf{0.022} & 0.090 & 0.090 & 0.093 & 0.94 & \textbf{-0.028} & 0.089 & 0.089 & 0.093 & 0.93\\

 & 100 & \textbf{0.023} & 0.049 & 0.048 & 0.053 & 0.93 & \textbf{-0.027} & 0.049 & 0.047 & 0.054 & 0.94\\

\multirow{-3}{*}{GEE (exch)} & 150 & \textbf{0.020} & 0.040 & 0.041 & 0.046 & \textbf{0.92} & \textbf{-0.025} & 0.040 & 0.040 & 0.047 & \textbf{0.90}\\
\cmidrule{1-12}
\multicolumn{12}{c}{Treatment 2} \\
\cmidrule(lr){1-12}
 & 30 & 0.002 & 0.091 & 0.090 & 0.090 & 0.96 & -0.001 & 0.092 & 0.089 & 0.089 & 0.96\\

 & 100 & 0.002 & 0.050 & 0.051 & 0.051 & 0.94 & -0.001 & 0.050 & 0.048 & 0.048 & 0.96\\

\multirow{-3}{*}{EMEE} & 150 & -0.003 & 0.041 & 0.041 & 0.041 & 0.94 & 0.002 & 0.041 & 0.039 & 0.039 & 0.96\\
\cmidrule{1-12}
 & 30 & 0.002 & 0.092 & 0.090 & 0.090 & 0.96 & -0.001 & 0.091 & 0.089 & 0.089 & 0.96\\

 & 100 & 0.002 & 0.050 & 0.051 & 0.051 & 0.95 & -0.001 & 0.050 & 0.048 & 0.048 & 0.96\\

\multirow{-3}{*}{EMEE-NonP} & 150 & -0.003 & 0.041 & 0.041 & 0.041 & 0.94 & 0.002 & 0.041 & 0.039 & 0.039 & 0.96\\
\cmidrule{1-12}
 & 30 & 0.008 & 0.085 & 0.090 & 0.091 & 0.93 & -0.008 & 0.077 & 0.088 & 0.088 & 0.92\\

 & 100 & 0.008 & 0.046 & 0.051 & 0.051 & 0.92 & -0.007 & 0.042 & 0.048 & 0.048 & 0.91\\

\multirow{-3}{*}{DR-EMEE-NonP} & 150 & 0.002 & 0.038 & 0.040 & 0.040 & 0.92 & -0.004 & 0.034 & 0.039 & 0.039 & 0.92\\
\cmidrule{1-12}
 & 30 & \textbf{0.019} & 0.086 & 0.089 & 0.091 & 0.94 & \textbf{-0.022} & 0.086 & 0.084 & 0.087 & 0.95\\

 & 100 & \textbf{0.019} & 0.047 & 0.049 & 0.053 & \textbf{0.92} & \textbf{-0.021} & 0.047 & 0.046 & 0.050 & 0.93\\

\multirow{-3}{*}{GEE (ind)} & 150 & \textbf{0.013} & 0.038 & 0.040 & 0.042 & \textbf{0.92} & \textbf{-0.019} & 0.038 & 0.037 & 0.042 & \textbf{0.92}\\
\cmidrule{1-12}
 & 30 & \textbf{0.019} & 0.086 & 0.089 & 0.091 & 0.94 & \textbf{-0.022} & 0.086 & 0.084 & 0.087 & 0.95\\

 & 100 & \textbf{0.018} & 0.047 & 0.050 & 0.053 & \textbf{0.92} & \textbf{-0.021} & 0.047 & 0.046 & 0.050 & 0.93\\

\multirow{-3}{*}{GEE (exch)} & 150 & \textbf{0.013} & 0.038 & 0.040 & 0.042 & \textbf{0.92} & \textbf{-0.019} & 0.038 & 0.037 & 0.042 & \textbf{0.92}\\
\bottomrule
\end{tabular}}
\label{simu8}
\end{table}

\end{appendices}
 
\end{document}